\documentclass[aps,prb,longbibliography,superscriptaddress,reprint]{revtex4-1}

\usepackage{graphicx,dcolumn,bm,amsmath,amssymb,euscript}

\newcommand{\VERSION}{February 15, 2021}  

\newcommand{\trel}{t_\mathrm{rel}}
\newcommand{\fass}{f_\mathrm{\alpha}^\mathrm{ss}}
\newcommand{\rhoeq}{\rho_\mathrm{eq}}
\newcommand{\Kss}{K^\mathrm{ss}}
\newcommand{\Kdyn}{K^\mathrm{dyn}}
\newcommand{\Gss}{G^\mathrm{ss}}
\newcommand{\Hm}{H_\mathrm{m}}
\newcommand{\Sm}{S_\mathrm{m}}
\newcommand{\muGa}{\mu_\mathrm{Ga}}
\newcommand{\yicalc}{y_i^\mathrm{calc}}

\usepackage[colorlinks,citecolor=red,urlcolor=blue]{hyperref}

\usepackage{color, colortbl}

\graphicspath{{figures/vmain/}}
	
\usepackage[update,prepend]{epstopdf}

\begin{document}

\title{\textit{In Situ} Microbeam Surface X-ray Scattering Reveals Alternating Step Kinetics During Crystal Growth}

\author{Guangxu Ju}
    \email[correspondence to: ]{juguangxu@gmail.com}
	\altaffiliation[current address: ]{Lumileds Lighting Co., San Jose, CA 95131 USA.}
	\affiliation{Materials Science Division, Argonne National Laboratory, Lemont, IL 60439 USA}
\author{Dongwei Xu}
	\affiliation{Materials Science Division, Argonne National Laboratory, Lemont, IL 60439 USA}
	\affiliation{School of Energy and Power Engineering, Huazhong University of Science and Technology, Wuhan, Hubei 430074, China}
\author{Carol Thompson}
	\affiliation{Department of Physics, Northern Illinois University, DeKalb, IL 60115 USA}
\author{Matthew J. Highland}
	\affiliation{X-ray Science Division, Argonne National Laboratory, Lemont, IL 60439 USA}
\author{Jeffrey A. Eastman}
	\affiliation{Materials Science Division, Argonne National Laboratory, Lemont, IL 60439 USA}
\author{Weronika Walkosz}
	\affiliation{Department of Physics, Lake Forest College, Lake Forest, IL 60045 USA}
\author{Peter Zapol}
	\affiliation{Materials Science Division, Argonne National Laboratory, Lemont, IL 60439 USA}
\author{G. Brian Stephenson}
    \email[correspondence to: ]{stephenson@anl.gov}
	\affiliation{Materials Science Division, Argonne National Laboratory, Lemont, IL 60439 USA}

\date{\VERSION}

\begin{abstract}
The stacking sequence of hexagonal close-packed and related crystals typically results in steps on vicinal $\{0001\}$ surfaces that have alternating $A$ and $B$ structures with different growth kinetics. However, because it is difficult to experimentally identify which step has the $A$ or $B$ structure, it has not been possible to determine which has faster adatom attachment kinetics. Here we show that \textit{in situ} microbeam surface X-ray scattering can {determine} whether $A$ or $B$ steps have faster kinetics under specific growth conditions. We demonstrate this for organo-metallic vapor phase epitaxy of (0001) GaN. X-ray measurements performed during growth find that the average width of terraces above $A$ steps increases with growth rate, indicating that attachment rate constants are higher for $A$ steps, in contrast to most predictions. Our results have direct implications for understanding the atomic-scale mechanisms of GaN growth {and can} be applied to a wide variety of related crystals. 
\end{abstract}

\maketitle


\section*{Introduction}
\begin{figure}[h]
\includegraphics[width=\linewidth]{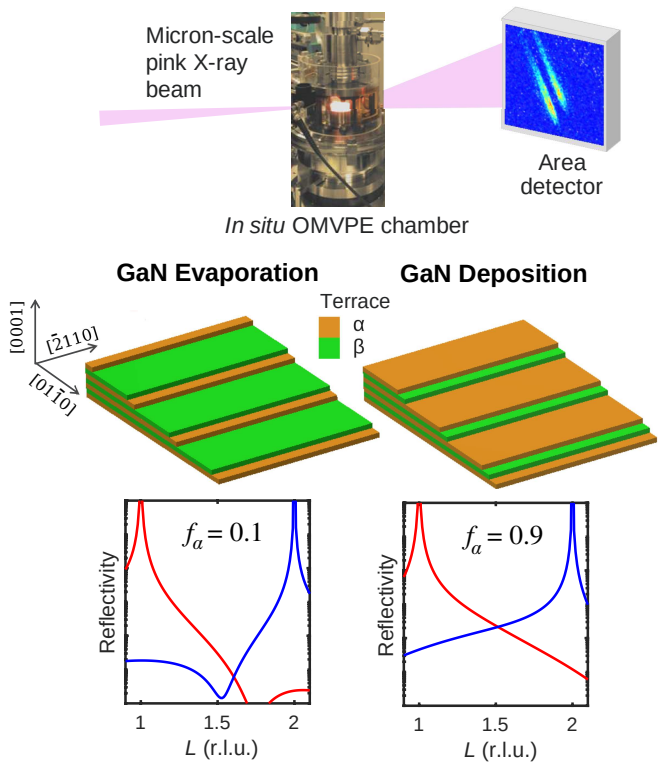} 
\caption{Schematic of microbeam surface X-ray scattering during organo-metallic vapor phase epitaxy (OMVPE) growth. Crystal truncation rod (CTR) measurements are sensitive to the $\alpha$ or $\beta$ terrace fraction changes that occur on vicinal $\{0001\}$ surfaces of HCP-type crystals during deposition or evaporation. Calculated reflectivities are shown for the CTRs from the $(0 1 \overline{1} 1)$ and $(0 1 \overline{1} 2)$ Bragg peaks (red and blue curves, respectively) with $\alpha$ terrace fractions $f_\alpha = 0.1$ and $0.9$  typical for evaporation and deposition of GaN by OMVPE, as will be shown below.
\label{fig:overview}}
\end{figure}

Our understanding of crystal growth is built on a powerful paradigm quantified by Burton, Cabrera, and Frank (BCF) \cite{1951_Burton_PhilTransRS243_29,1999_Jeong_SurfSciRep34_171,2015_Woodruff_PhilTransA373_20140230}, in which atoms are added to the growing crystal surface by attachment at the steps forming the edges of each exposed atomic layer, or terrace.
The BCF model was originally developed for crystals with step heights of a full unit cell and step properties that are identical from step to step for a given step direction.
However, from the beginning \cite{1951_Verma_PhilMag42_1005} it was recognized that there could be more complex situations.
When the space group of the crystal includes screw axes or glide planes,
the growth behavior can be fundamentally different on facets perpendicular to one of these symmetry elements \cite{2004_vanEnckevort_ActaCrystA60_532}.
In this case, the terraces can still all have the same atomic arrangement, but now have different in-plane orientations of their top layer.
The fractional-unit-cell-height steps that separate these terraces have structures and properties that can vary from step to step, even for a fixed step direction.
Thus surface morphologies with alternating terrace widths can arise that depend upon the deposition or evaporation conditions, as indicated in Fig.~\ref{fig:overview}.
The inequivalent kinetics at steps affects not only surface morphology but also the incorporation of alloying elements during crystal growth \cite{2013_Turski_JCrystGrowth367_115,2016_Kaufmann_JCrystGrowth433_36}.

\begin{figure}[b]
\includegraphics[width=\columnwidth]{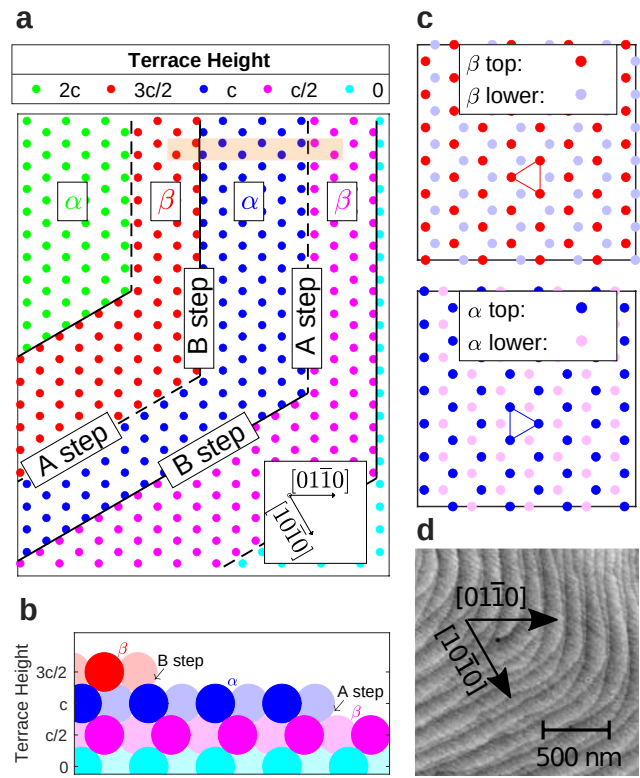}
\caption{Terrace and step structure of vicinal (0001) surface of an HCP-type crystal.  For GaN, only Ga atoms are shown. \textbf{a}. Circles show top-layer sites on each terrace, with color indicating height. Steps typically have lowest edge energy when they are normal to $[0 1 \overline{1} 0]$, $[1 0 \overline{1} 0]$, or $[1 \overline{1} 0 0]$. Steps in a sequence have alternating structures, $A$ and $B$, which swap when step azimuth changes by $60^\circ$. \textbf{b}. Cross-section of the $A$ and $B$ step structures in the region marked by an orange rectangle in \textbf{a}. Lighter and darker colors indicate atoms in different rows.  \textbf{c}. Detail of $\alpha$ and $\beta$ terrace structures. Orientation of triangle of top-layer atoms around $6_3$ screw axis shows difference between layers.  \textbf{d}. AFM height image of GaN (0001) surface typical of films grown on sapphire substrates by OMVPE, showing regions of alternating step spacings and interlacing at corners where the step azimuth changes. Step heights are $c/2 = 2.6$~\AA.
\label{fig:A_B}}
\end{figure}

A ubiquitous but subtle version of this effect occurs on the basal-plane $\{0001\}$-type surfaces of crystals having hexagonal close-packed (HCP) or related structures, which are normal to a $6_3$ screw axis.
Such crystals are made up of close-packed layers with 3-fold symmetry that alternate between opposite orientations, as shown by the $\alpha$ and $\beta$ terrace structures in Fig.~\ref{fig:A_B}(c).
On a vicinal surface, the $\alpha \beta \alpha \beta$ stacking sequence typically results in half-unit-cell-height steps.
The lowest energy steps are normal to $[0 1 \overline{1} 0]$-type directions, and have alternating structures conventionally labelled $A$ and $B$ \cite{1999_Xie_PRL82_2749,2001_Giesen_ProgSurfSci68_1}
as shown in Fig.~\ref{fig:A_B}(a,b).
When the in-plane azimuth of a step changes by $60^\circ$, e.g. from $[0 1 \overline{1} 0]$ to $[1 0 \overline{1} 0]$, its structure changes from $A$ to $B$ or $B$ to $A$.

The alternating nature of the steps on such surfaces has been imaged in several systems, including SiC \cite{1951_Verma_PhilMag42_1005}, GaN \cite{1999_Xie_PRL82_2749,1999_Heying_JAP85_6470,2001_Vezian_MatSciEngB82_56,2006_Xie_PRB_085314,2008_Zheng_PRB77_045303,2013_Turski_JCrystGrowth367_115}, AlN \cite{2017_Pristovsek_physstatsolb254_1600711}, and ZnO \cite{2002_Chen_APL80_1358}. 
These systems typically show a tendency for local pairing of steps (i.e. alternating step spacings), and an interlaced structure in which the step pairs switch partners at corners where their azimuth changes by $60^\circ$, as shown in Fig.~\ref{fig:A_B} (d).
These features are consistent with predictions that $A$ and $B$ steps have significantly different attachment kinetics \cite{1999_Xie_PRL82_2749,2006_Xie_PRB_085314,2013_Turski_JCrystGrowth367_115,2010_Zaluska-Kotur_JNoncrystSolids356_1935,2011_Zaluska-Kotur_JAP109_023515,2017_Chugh_ApplSurfSci422_1120, 2017_Xu_JChemPhys146_144702,2020_Akiyama_JCrystGrowth532_125410,2020_Akiyama_JJAP59_SGGK03,ohka2020effect} that lead to unequal local fractions of $\alpha$ and $\beta$ terraces during growth.
However, it has not been possible to experimentally distinguish the terrace orientation or step structure, and thus to determine whether $A$ or $B$ steps have faster kinetics.

{In particular, the properties of $A$ and $B$ steps on GaN (0001) surfaces have been a matter of some disagreement.}
A seminal study \cite{1999_Xie_PRL82_2749} of molecular beam epitaxy (MBE) of GaN observed alternating step shapes and proposed that the kinetic coefficients for adatom attachment are higher for $A$ steps than $B$ steps, i.e. $A$ steps grow faster for a given supersaturation.
The support for this highly cited prediction is based on an argument regarding the difference in dangling bonds between $A$ and $B$ steps, and a comparison with experimental results on GaAs (111) \cite{1997_Avery_PRL79_3938,2012_Jo_CrystGrowthDes12_1411}. 
(Such face-centered cubic materials have $A$ and $B$ type steps that do not alternate between successive terraces and thus can be distinguished by their orientation \cite{2001_Giesen_ProgSurfSci68_1}.)
In contrast, several subsequent theoretical studies of GaN (0001) {organo-metallic vapor phase epitaxy (OMVPE)} and MBE have consistently predicted that $A$ steps have smaller adatom attachment coefficients than $B$ steps.
Kinetic Monte Carlo (KMC) studies of GaN (0001) growth under OMVPE conditions found step pairing \cite{2011_Zaluska-Kotur_JAP109_023515} driven by faster kinetics at $B$ steps than $A$ steps \cite{2010_Zaluska-Kotur_JNoncrystSolids356_1935}.
The standard bond-counting energetics used in a KMC study of growth on an HCP lattice \cite{2017_Xu_JChemPhys146_144702} result in a much lower Ehrlich-Schwoebel (ES) barrier at $B$ steps than at $A$ steps, when only nearest-neighbor jumps are allowed.
A recent KMC study of GaN (0001) growth under MBE conditions \cite{2017_Chugh_ApplSurfSci422_1120} found triangular islands that close analysis reveals are bounded by $A$ steps, indicating faster growth of $B$ steps.
An analysis of InGaN (0001) growth by MBE \cite{2013_Turski_JCrystGrowth367_115} concluded that adatom attachment at $B$ steps is faster, converting them into crenelated edges terminated by $A$ steps.

The difference between the kinetics at $A$ and $B$ steps is a reflection of the chemical states of the adatoms, steps and terraces that affect the dynamics of $A$ and $B$ steps.
Studies of islands on the FCC Pt (111) surface \cite{1998_Kalff_PRL81_1255,2009_Yin_APL94_183107} have found that $A$ steps have a higher growth rate than $B$ steps, but that this relationship is reversed by the presence of adsorbates such as CO.
An experimental study of AlN (0001) surfaces grown by OMVPE \cite{2017_Pristovsek_physstatsolb254_1600711} found a change in the terrace fraction as a function of the V/III ratio used during growth.
\textit{Ab initio} calculations of kinetic barriers on GaN and AlN (0001) under MBE and OMVPE conditions \cite{2020_Akiyama_JCrystGrowth532_125410,2020_Akiyama_JJAP59_SGGK03,ohka2020effect} found that the barriers and adsorption energies at $A$ and $B$ steps depend in detail on the surface reconstruction induced by the environment.
Thus to properly understand, model, and control (0001) surface morphology in HCP-type systems, there is a need for an \textit{in situ} experimental method that can distinguish adatom attachment kinetics at $A$ and $B$ steps in the relevant growth environment.

{Here we show that \textit{in situ} surface X-ray scattering can distinguish the fraction of the surface covered by $\alpha$ or $\beta$ terraces during growth, unambiguously determining differences in the attachment kinetics at $A$ and $B$ steps.
X-rays are an ideal probe since they are sensitive to atomic-scale structure and can penetrate the growth environment.
This method is enabled by using a micron-scale X-ray beam that illuminates a surface region of a high-quality single crystal having a uniform step azimuth, as shown in Fig.~\ref{fig:overview}.
We demonstrate this for OMVPE of (0001) GaN, with measurements of crystal truncation rods (CTRs) carried out \textit{in situ} during growth.}
CTRs are streaks of intensity extending in reciprocal space away from every Bragg peak in the direction normal to the crystal surface, which are sensitive to the surface structure \cite{1986_Robinson_PRB33_3830}.
We fit calculated CTRs from a model structure to these measurements to obtain the variation of the steady-state $\alpha$ terrace fraction $f_\alpha$ as a function of growth conditions, as well as the relaxation times $\trel$ of $f_\alpha$ upon changing conditions.
These results are compared to calculated dynamics based on a BCF model for a system with alternating step types, to quantify the differences in the attachment rates at $A$ and $B$ steps.

\section*{Results}

\textbf{Calculated surface X-ray scattering with alternating step types.} We present calculations showing how the intensity distribution along the CTRs is sensitive to the the fraction of the surface covered by $\alpha$ or $\beta$ terraces.
Our calculations include the effect of surface reconstruction, using relaxed atomic coordinates that have been obtained previously \cite{2012_Walkosz_PRB_85_033308}.
For a vicinal surface, the CTRs are tilted away from the crystal axes, so that the CTRs from different Bragg peaks do not overlap.
The X-ray reflectivity along the CTRs can be calculated by adding the complex amplitudes from the substrate crystal and the reconstructed overlayers, with proper phase relationships
\cite{1999_Munkholm_JApplCryst32_143,2002_Trainor_JApplCryst35_696,2017_Petach_PRB95_184104}.
Details of our calculations are given in a separate paper \cite{2020_Ju_PRB_CTR_arxiv}.

\begin{figure}[h]
\includegraphics[width=\columnwidth]{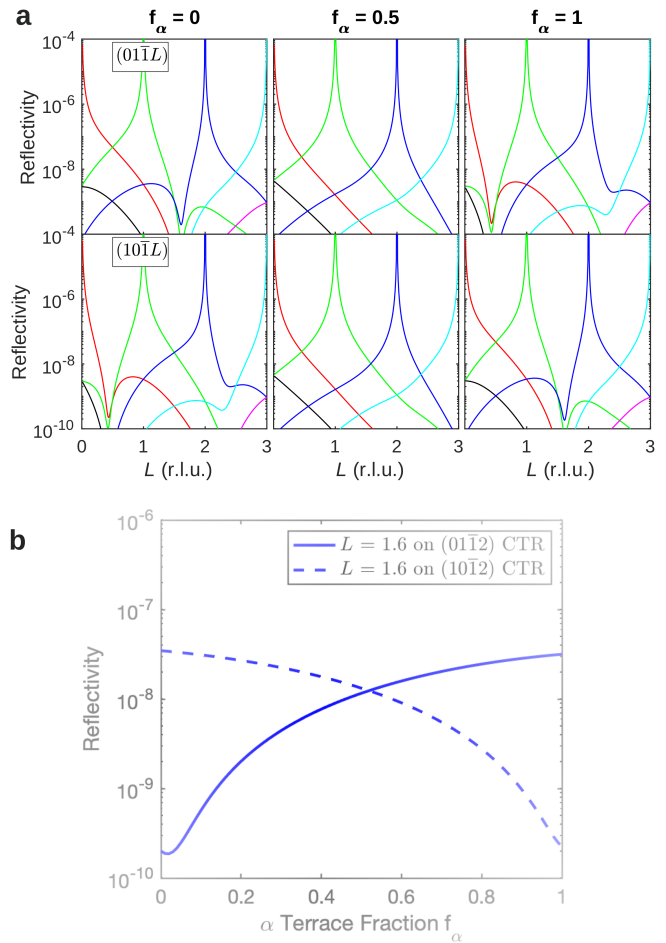} 
 \caption{ Calculated CTR intensities for a vicinal GaN $(0 0 0 1)$ surface. \textbf{a}. Top and bottom rows show the families of CTRs along $(0 1 \overline{1} L)$ and $(1 0 \overline{1} L)$, respectively. Black, red, green, blue, cyan, and magenta curves are for CTRs from the $L_0 = -1$ to $4$ Bragg peaks, respectively. Values of $f_\alpha$ for each column are given at the top. \textbf{b}. Reflectivity of selected CTRs as a function of terrace fraction $f_\alpha$ for fixed values of $L = 1.6$. \label{fig:CTR_miscut_3HT1} }
\end{figure}

Figure \ref{fig:CTR_miscut_3HT1}(a) shows calculated intensity distributions along $(0 1 \overline{1} L)$ and $(1 0 \overline{1} L)$ CTRs for the GaN $(0 0 0 1)$ surface, demonstrating how their $L$-dependences vary with $f_\alpha$. Bragg peak locations have integer indices $H_0 K_0 L_0$ in reciprocal lattice units; these indices identify the CTR associated with each peak. 
For this comparison, we use the 3H(T1) surface reconstruction\cite{2012_Walkosz_PRB_85_033308} and a fixed surface roughness, independent of $f_\alpha$,
as discussed in the Methods section below and Supplementary Discussion 1.
The same qualitative behavior is obtained using other surface reconstructions. 
For $f_\alpha = 0$ and $f_\alpha = 1$, the regions between the Bragg peaks have alternating stronger and weaker intensities, with the alternation being opposite for $(0 1 \overline{1} L)$ and $(1 0 \overline{1} L)$.
For $f_\alpha = 0.5$, the intensities between all Bragg peaks are about the same, and there is no difference between the $(0 1 \overline{1} L)$ and $(1 0 \overline{1} L)$ CTRs.
As required by symmetry, the $(0 1 \overline{1} L)$ CTRs with $f_\alpha = X$ are identical to the $(1 0 \overline{1} L)$ CTRs with $f_\alpha = 1-X$, for any value $X$.
Fig.~\ref{fig:CTR_miscut_3HT1}(b) shows calculations of the reflectivity as a function of $f_\alpha$ at positions near $L = 1.6$ on the $(0 1 \overline{1} 2)$ and $(1 0 \overline{1} 2)$ CTRs. The variation in reflectivity is almost monotonic in $f_\alpha$ at these positions.
These curves are used below to extract $f_\alpha(t)$ during dynamic transitions.

\begin{table*}[t]
\caption{ \label{tab:table1} Growth conditions studied. Also given are values of steady-state terrace fraction $\fass$ obtained from fits to measured CTR intensities (shown in Fig.~\ref{fig:fit_nonspecular_exp_3HT1}),
values of  relaxation time $\trel$ of $f_\alpha$ upon changing conditions (shown in Fig.~\ref{fig:relaxation_time}), and corresponding values from the BCF model fit (shown in Fig.~\ref{fig:fss_exp}).  } 
\begin{ruledtabular}
\begin{tabular}{ c  c  c  c || c  c  c}   
Growth & TEGa	flow & H$_2$ frac. &	Net growth & & Measured Values& Best Fit\\
condition & ($\mu$mole & in &	rate $G$ & & from & from\\
index &  min$^{-1}$) & carrier &	(monolayer s$^{-1}$) & & CTR fits & BCF model\\
\hline
1&0.000 & 50\% & -0.0018 & $\fass$ & $0.111 \pm 0.013$ & 0.136\\
2 & 0.000 & 0\% & 0.0000 & $\fass$ & $0.461 \pm 0.018$ & 0.440\\
3 & 0.033 & 50\% & 0.0109 & $\fass$ & $0.811 \pm 0.014$ & 0.836\\
4 & 0.033 & 0\% & 0.0127 & $\fass$ & $0.868 \pm 0.011$ & 0.847\\
\hline
$1$ to $2$ & & & & $\trel$ & $2200 \pm 200$ s & $2478$ \\
$2$ to $4$ & & & & $\trel$ & $340 \pm 30$ s & $331$ \\
\end{tabular}
\end{ruledtabular}
\end{table*}

\begin{figure}[h]
\includegraphics[width=0.8\columnwidth]{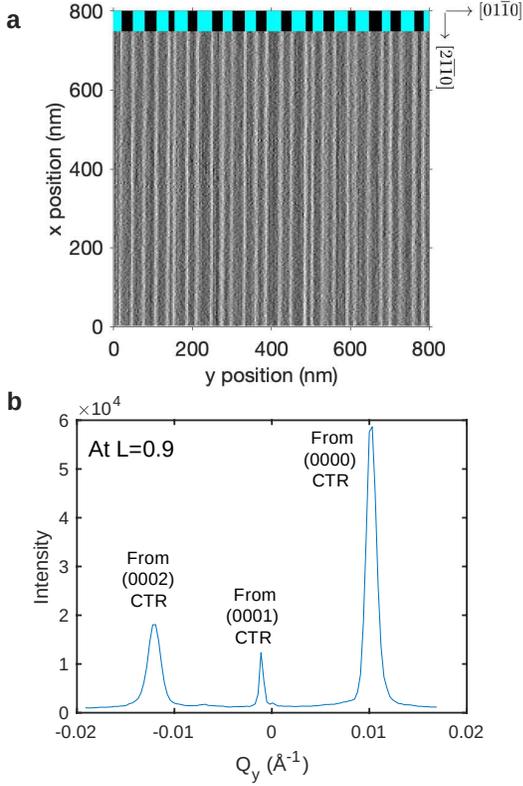}
\caption{\label{fig:AFM} {Imaging and scattering from steps.}  \textbf{a}. AFM image of steps of height c/2 (2.6 \AA) on the vicinal GaN substrate used in X-ray measurements. To emphasize the positions of steps, we plot the amplitude error signal, which is proportional to the height gradient in the scan direction $(y)$. Image was obtained \textit{ex situ} at room $T$ after an anneal for 300s at 1118 K in zero-growth conditions (0\% H$_2$, 0 TEGa). Average fraction over a $2 \times 2$~$\mu$m area of terrace family marked black at top is 0.47. The average double-step spacing of $w = 573$~\AA~corresponds to a miscut angle of $\tan^{-1}(c/w) = 0.52^\circ$. 
\textbf{b}. Profile of split CTRs from vicinal surface, measured at $T = 1170$~K during growth at $0.053$~$\mu$mole min$^{-1}$ TEGa, 50\% H$_2$. The splitting of the CTRs $\Delta Q_y = 0.0110$~\AA$^{-1}$ corresponds to a miscut angle of $\tan^{-1}[\Delta Q_y / (2 \pi / c)] = 0.52^\circ$, in agreement with the AFM result.}
\end{figure}

\textbf{\textit{In situ} X-ray scattering measurements during growth.} We studied four OMVPE conditions having different net growth rates at the same temperature $1076 \pm 5$~K, summarized in Table~\ref{tab:table1} (see Methods for further details).
The substrate used was a GaN single crystal. 
Fig.~\ref{fig:AFM}(a) shows its initial surface morphology determined by \textit{ex situ} atomic force microscopy (AFM).
One can see straight steps almost perpendicular to the $y$ or $[0 1 \overline{1} 0]$ direction over large areas.
An analysis of the step spacing shows a slight tendency towards pairing, with one of the two alternating terrace types having an area fraction of 0.47.
AFM is insensitive to whether this fraction corresponds to the $\alpha$ or $\beta$ terraces.
We also characterized the miscut angle by measuring the splitting of the CTRs.
Fig.~\ref{fig:AFM}(b) shows a transverse cut through the CTRs in the $Q_y$ direction near $(0 0 0 L)$ at $L = 0.9$.
Both the AFM and X-ray measurements give a double-step spacing of $w = 573$~\AA~ corresponding to a miscut angle of $0.52^\circ$.
To relate the $\alpha$ terrace fraction to the behavior of $A$ and $B$ steps, it is critical to determine the sign of the step azimuth.
By making measurements as a function of $L$, we verified that the peak at high $Q_y$ is the CTR coming from $(0 0 0 0)$, while the peak at low $Q_y$ is the $(0 0 0 2)$ CTR.
This confirms that the downstairs direction of the vicinal surface is in the $+y$ direction, as drawn in Fig.~\ref{fig:overview}.
It is also useful to know the precise angle of the step azimuth with respect to the crystal planes, which determines the minimum kink density and thus the predicted values of some kinetic coefficients.
X-ray measurements found this to be $5^\circ$ off of the $[0 1 \overline{1} 0]$ direction towards $[1 0 \overline{1} 0]$, which gives a maximum kink spacing of 33~\AA. 
The kink spacing could be smaller due to thermally generated kinks \cite{1951_Burton_PhilTransRS243_29}, {as discussed in Supplementary Discussion 2}.
With this low-dislocation-density substrate and the low growth rates used, the previously reported instability to step bunching during growth \cite{2000_Murty_PRB62_R10661} was not observed.

\begin{figure}[b]
\includegraphics[width=\linewidth]{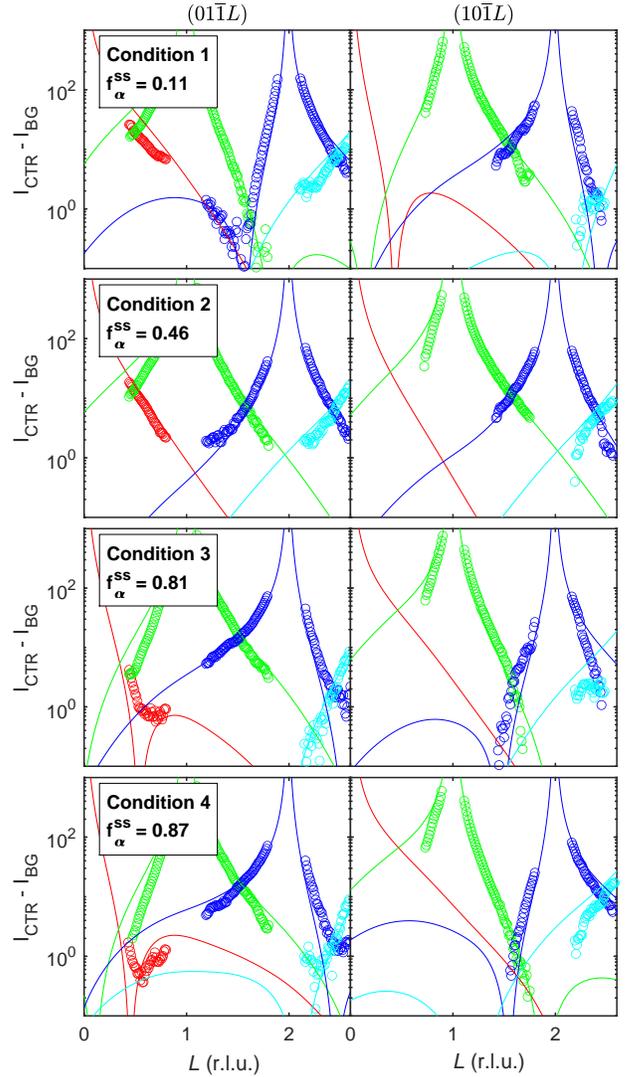}
\caption{{Measured and calculated crystal truncation rods.} Symbols show measured net intensities of the $(0 1 \overline{1} L_0)$ CTRs and the $(1 0 \overline{1} L_0)$ CTRs families (left and right) for CTRs from the $L_0 = 0,1,2,3$ Bragg peaks at each of four conditions. Curves show fits of all CTRs to obtain steady-state $\alpha$ terrace fraction $\fass$ at each condition.
\label{fig:fit_nonspecular_exp_3HT1}}
\end{figure}

Figure~\ref{fig:fit_nonspecular_exp_3HT1} shows the measured steady-state CTR intensities as a function of $L$, for both the $(0 1 \overline{1} L)$ and $(1 0 \overline{1} L)$ CTRs and at all four conditions.
The qualitative behavior agrees with that expected from a variation in $f_\alpha$ shown in Fig.~\ref{fig:CTR_miscut_3HT1}(a), with alternating higher and lower intensities between the Bragg peaks under some conditions, and opposite behavior of the two CTRs. 

\textbf{Steady-state and dynamics of terrace fraction during growth.} To obtain values of the steady-state terrace fraction $\fass$ for each of the four conditions, we fit calculated CTR intensities to the measured profiles. 
We performed fits using different possible surface reconstructions (see Methods for further details). 
While the 3H(T1) reconstruction gives the best fit to all conditions, similar $\fass$ values are obtained using alternative reconstructions. 
For each condition, both the $(0 1 \overline{1} L)$ and $(1 0 \overline{1} L)$ CTRs were simultaneously fit.
The values of $\fass$ obtained as a function of net growth rate $G$ are given in Table~\ref{tab:table1}.  
The marked increase in $\fass$ as $G$ is increased reveals the qualitative difference between the kinetics at $A$ and $B$ steps during OMVPE of GaN: adatom attachment coefficients for $A$ steps are larger.
Thus a surface with initially balanced $\alpha$ and $\beta$ terrace fractions at zero net growth rate will evolve to one with higher $\fass$ during positive net growth, because of the initially higher adatom attachment rate at the $A$ steps.
Likewise, during evaporation the initially higher detachment rate at $A$ steps will give a lower $\fass$.

\begin{figure}[b]
\includegraphics[width=0.9\linewidth]{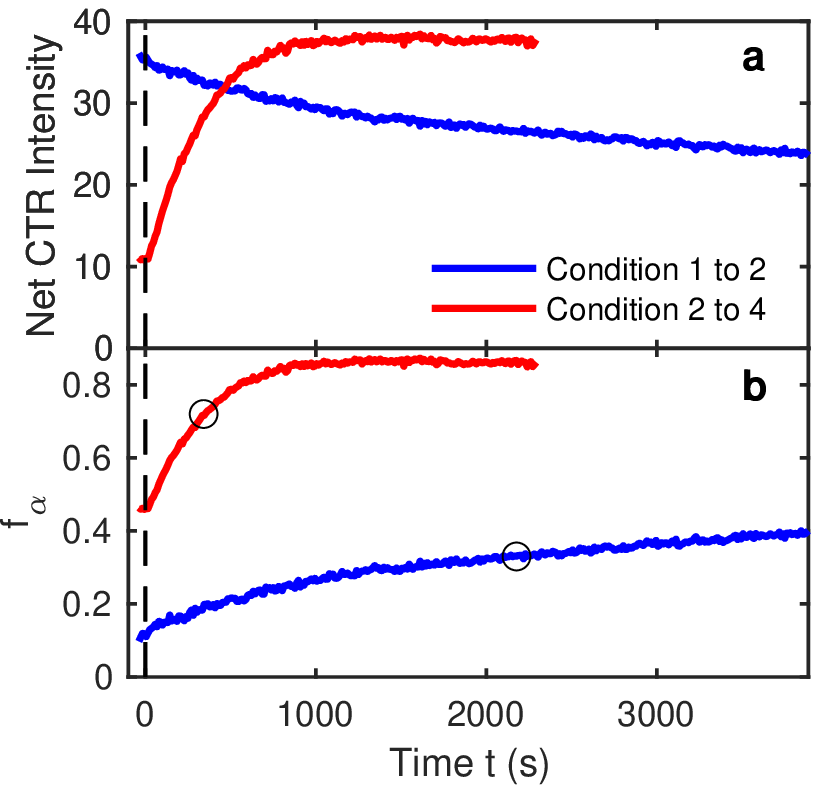}
\caption{Dynamics following a change of condition at $t = 0$. 
\textbf{a}. Measured CTR intensities.
\textbf{b}. Calculated terrace fractions $f_\alpha$.
Blue curves: condition 1 to 2 on the $(1 0 \overline{1} 2)$ CTR at $L=1.627$. Red curves: condition 2 to 4 on the $(0 1 \overline{1} 2)$ CTR at $L=1.603$. Circles show $1/e$ relaxation times $\trel$ of 2200 $\pm$ 200 s and 340 $\pm$ 30 s, respectively. 
\label{fig:relaxation_time}}
\end{figure}

We also observed the dynamics of the change in $f_\alpha$ by recording the intensity at a fixed detector position as a function of time before and after an abrupt change between conditions, as shown in
Fig.~\ref{fig:relaxation_time}(a). 
We chose positions near $L = 1.6$ where the X-ray reflectivity $R$ changes almost monotonically with $f_\alpha$, as shown in Fig.~\ref{fig:CTR_miscut_3HT1}(b).
It is thus straightforward to obtain $f_\alpha(t)$ from the intensity evolution using the calculated $R(f_\alpha)$,
as shown in Fig.~\ref{fig:relaxation_time}(b).
The characteristic $1/e$ relaxation times $\trel$ were $2200 \pm 200$ s and $340 \pm 30$ s for the transitions from conditions 1 to 2 and 2 to 4, respectively.

\begin{figure}[b]
\includegraphics[width=0.9\linewidth]{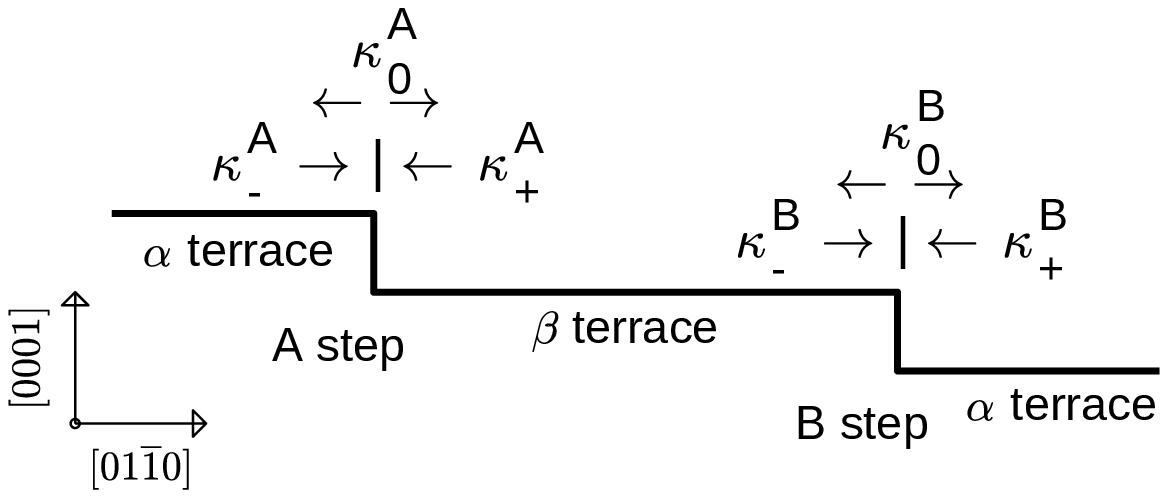}
\caption{Schematic of alternating terraces and steps of BCF model for HCP basal plane surfaces, showing kinetic coefficients for the $A$ and $B$ steps. \label{fig:BCF}}
\end{figure}

\textbf{BCF model for surface with alternating step types.} To quantitatively relate the behavior of the terrace fraction to the kinetic properties of $A$ and $B$ steps, we have developed a model \cite{2020_Ju_PRB_BCF} based on BCF theory.
Such models have been used extensively to understand growth behavior such as the step-bunching instability \cite{2020_Guin_PRL124_036101}, pairing of steps \cite{2004_Pierre-Louis_PRL93_165901}, and competitive adsorption \cite{2019_Hanada_PhysRevMater3_103404}, typically where all steps in a sequence are equivalent.
In our model, we consider an alternating sequence of two types of terraces, $\alpha$ and $\beta$, and two types of steps, $A$ and $B$, with properties that can differ, as shown in Fig.~\ref{fig:BCF}.
Related BCF models with alternating step or terrace properties have appeared previously \cite{zhao2015refined,2011_Zaluska-Kotur_JAP109_023515, 2010_Zaluska-Kotur_JNoncrystSolids356_1935,2007_Sato_EurPhysJB59_311,2006_Xie_PRB_085314,2005_Frisch_PRL94_226102}.
We include the effects of step transparency \cite{2003_Pierre-Louis_PRE68_021604} (i.e. adatom transmission across steps) and step-step repulsion \cite{1999_Jeong_SurfSciRep34_171}.

The rate of change in the adatom density per unit area $\rho_i$ on terrace type $i = \alpha$ or $\beta$ is written as
\begin{equation}
    \frac{\partial \rho_i}{\partial t} = D \nabla^2 \rho_i - \frac{\rho_i}{\tau} + F,
    \label{eq:cont}
\end{equation}
where $D$ is the adatom diffusivity, $\tau$ is the adatom lifetime before evaporation, and $F$ is the deposition flux of adatoms per unit time and area. 
The four boundary conditions for the flux at the steps terminating each type of terrace can be written as
\begin{align}
    - D\nabla \rho_\alpha^+
    &= +\kappa_-^A (\rho_\alpha^+ - \rhoeq^A)
    + \kappa_0^A (\rho_\alpha^+ - \rho_\beta^-), \label{eq:bc1} \\
    - D\nabla \rho_\alpha^-
    &= -\kappa_+^B (\rho_\alpha^- - \rhoeq^B)
    - \kappa_0^B (\rho_\alpha^- - \rho_\beta^+), \label{eq:bc2} \\
    - D\nabla \rho_\beta^+
    &= +\kappa_-^B (\rho_\beta^+ - \rhoeq^B)
    + \kappa_0^B (\rho_\beta^+ - \rho_\alpha^-), \label{eq:bc3} \\
    - D\nabla \rho_\beta^-
    &= -\kappa_+^A (\rho_\beta^- - \rhoeq^A)
    - \kappa_0^A (\rho_\beta^- - \rho_\alpha^+). \label{eq:bc4}
\end{align}
As shown in Fig.~\ref{fig:BCF}, $\kappa_+^j$ and $\kappa_-^j$ are the kinetic coefficients for adatom attachment at a step of type $j = A$ or $B$ from below or above, respectively,
and $\kappa_0^j$ is the kinetic coefficient for transmission across the step.
The $+$ or $-$ superscripts on $\rho_i$ and $\nabla \rho_i$ indicate evaluation at the downhill or uphill terrace boundaries, respectively. 
We consider the overall vicinal angle of the surface to fix the sum $w$ of the widths of $\alpha$ and $\beta$ terraces, which are thus $f_\alpha w$ and $(1-f_\alpha) w$.
We also assume relations between the equilibrium adatom densities $\rhoeq^j$ at the steps and the terrace widths that reflect an effective repulsion between the steps owing to entropic and strain effects \cite{1999_Jeong_SurfSciRep34_171},
\begin{equation}
    \rhoeq^j = \rhoeq^0 \exp(\mu_j/kT),
\end{equation}
where $\rhoeq^0$ is the equilibrium adatom density at zero growth rate, and the adatom chemical potentials $\mu_j$ at steps of type $j = A$ or $B$ have a dependence on $f_\alpha$ given by
\begin{equation}
    \frac{\mu_A}{kT} = -\frac{\mu_B}{kT} = M(f_\alpha) \equiv \left ( \frac{\ell}{w} \right )^3 \left [ \left ( \frac{1 - f_\alpha^0}{1 - f_\alpha} \right )^3 - \left ( \frac{f_\alpha^0}{f_\alpha} \right)^3 \right ],
\end{equation}
where $\ell$ is the step repulsion length and $f_\alpha^0 $ is the terrace fraction at zero growth rate.

To solve this model we develop a quasi-steady-state expression for the dynamics of the terrace fraction $f_\alpha$.
Under fairly general assumptions \cite{2020_Ju_PRB_BCF}, the behavior of $f_\alpha$ can be written as a function of the net growth rate,
\begin{equation}
    G = \frac{F - \rhoeq^0 / \tau}{\rho_0}.  \label{eq:Ga}
\end{equation}
This is simply the difference between the deposition $F$ and a uniform evaporation $\rhoeq^0 / \tau$, converted to monolayer per second (ML s$^{-1}$) using the site density $\rho_0$ per half-unit-cell-thick monolayer.
The rate of change in $f_\alpha$ is 
\begin{equation}
    \frac{df_\alpha}{dt} = \Kdyn(f_\alpha) \left ( \frac{G}{\Kss(f_\alpha)} - \frac{4 M(f_\alpha) \rhoeq^0}{w \rho_0} \right ), \label{eq:dfdta}
\end{equation}
where we have introduced the net steady-state and dynamic kinetic coefficient functions $\Kss(f_\alpha)$ and $\Kdyn(f_\alpha)$, which in the general case depend on all six $\kappa_x^j$ coefficients \cite{2020_Ju_PRB_BCF}.

The full steady state $df_\alpha/dt = 0$ is obtained at a growth rate of
\begin{equation}
    \Gss(f_\alpha) = \frac{4 \, \Kss(f_\alpha) \, M(f_\alpha) \rhoeq^0 }{w \rho_0}.
    \label{eq:fSSG}
\end{equation}
This equation for $\Gss(f_\alpha)$ can be inverted to obtain the steady-state value $\fass$ as a function of $G$.
The sign of $d\Gss/df_\alpha$ and thus $d\fass/dG$ is determined by the sign of $\Kss$.
General expressions for $\Kss$ show that in most cases, such as that found here, its sign is positive when the kinetic coefficients for $A$ steps are larger than those for $B$ steps \cite{2020_Ju_PRB_BCF}.

\begin{figure}[h]
\includegraphics[width=0.8\linewidth]{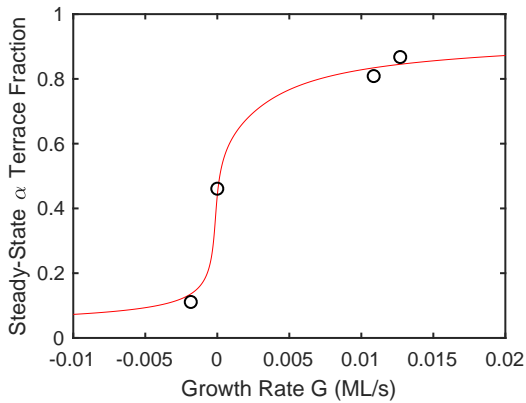}
\caption{Steady-state $\alpha$ terrace fraction $\fass$ as a function of growth rate $G$. Circles are experimental values given in Table~\ref{tab:table1}, showing monotonic increase of $\fass$ with increasing $G$. Curve is best-fit BCF model calculation; also fit simultaneously with these four points were the two relaxation times $\trel$ given in Table~\ref{tab:table1}. 
\label{fig:fss_exp}}
\end{figure}

To calculate BCF model results to compare with the experimental conditions, we make 3 assumptions: 1) the only parameter affected by the TEGa supply rate is the deposition flux $F$; 2) the only parameter affected by the carrier gas composition (0\% or 50\% H$_2$) is the adatom lifetime $\tau$; and 3) $F$ and $\tau$ enter only through the net growth rate $G$ given by Eq.~(\ref{eq:Ga}), listed in Table \ref{tab:table1} for each condition.
We use the known values $\rho_0 = 2 a^{-2} / \sqrt{3} = 1.13 \times 10^{19}$~m$^{-2}$ and $w = c / \sin(0.52^\circ) = 5.73 \times 10^{-8}$~m, where $a = 3.20 \times 10^{-10}$~m and $c = 5.20 \times 10^{-10}$~m are the lattice parameters of GaN at the growth temperature \cite{2000_Reeber_JMaterRes15_40}.
We performed fits to the measured quantities (four $\fass$ and two $\trel$) using the general expressions for $\Kss(f_\alpha)$ and $\Kdyn(f_\alpha)$ \cite{2020_Ju_PRB_BCF}. 
The relaxation times for the model were calculated by integrating Eq.~(\ref{eq:dfdta}) to obtain $f_\alpha(t)$, and then extracting the relaxation time with the same normalization procedure used for the experimental data. 
The best fit, shown in Fig.~\ref{fig:fss_exp}, was obtained in the limit in which the $\kappa_+^A$ coefficient is large, while the $\kappa_-^A$, $\kappa_-^B$, and $\kappa_0^A$ coefficients approach zero.
In this case $\Kss(f_\alpha)$ and $\Kdyn(f_\alpha)$ can be written as \cite{2020_Ju_PRB_BCF}
\begin{align}
    \Kss(f_\alpha) &= \left [ 
    \frac{1}{\kappa_+^B} + \frac{(1 - 2 f_\alpha)}{\kappa_0^B} - \frac{w f_\alpha (1-f_\alpha)}{D} \right ]^{-1},
    \label{eq:Dkm} \\
    \Kdyn(f_\alpha) &=
    \left [ \frac{1}{\kappa_+^B} + \frac{1}{\kappa_0^B} + \frac{w(1-f_\alpha)}{D} \right ]^{-1} . \label{eq:dfdtm}
\end{align}
Note that $\kappa_+^A$ does not appear in these expressions because in this limit other microscopic processes are rate-limiting. 
We can fit the measurements directly using these expressions to obtain the four parameters $D / \kappa_+^B$ = $1.9 \times 10^{-8}$~m, $D / \kappa_0^B$= $1.1 \times 10^{-8}$~m, $D \rhoeq^0 \ell^3$ = $3.3 \times 10^{-23}$~m$^3$~s$^{-1}$, and $f_\alpha^0$ = $0.44$. Table~\ref{tab:table1} compares the six measured quantities (four $\fass$ and two $\trel$) to the best-fit values calculated from the BCF model.

\section*{Discussion}

To interpret the combined parameters obtained from the fits, it is useful to estimate the adatom diffusivity $D$ and equilibrium adatom density $\rhoeq^0$.
\textit{Ab initio} calculations of the activation energy for Ga diffusion on the Ga-terminated (0001) surface have given values of $\Delta \Hm = 0.4$~eV \cite{1998_Zywietz_APL73_487} and $\Delta \Hm = 0.5$~eV \cite{2017_Chugh_PCCP19_2111}, and similar values have been obtained for 3d transition metal adatoms \cite{2011_Gonzalez-Hernandez_JAP110_083712}.
If we estimate the diffusivity from the \textit{ab initio} calculations using $D = a^2 \nu \exp(\Delta \Sm / k) \exp(-\Delta \Hm / kT)$ \cite{1989_Shewmon_2ndEd_DiffusioninSolids}, with $a = 3.2 \times 10^{-10}$~m, $\nu = 10^{13}$~s$^{-1}$, $\Delta \Sm = 0$, and $\Delta \Hm = 0.4$~eV, we obtain $D = 1.4 \times 10^{-8}$~m$^2$~s$^{-1}$ at $T = 1073$~K.
An analysis of spatial correlations in surface morphology of GaN films \cite{2014_Koleske_JCrystGrowth391_85} indicated a cross-over at $T = 1073$~K from surface diffusion transport to evaporation/condensation transport at a length scale of $\lambda = 1.5 \times 10^{-6}$~m for OMVPE growth with H$_2$ present in the carrier gas.
Thus the adatom lifetime $\tau$ can be estimated as $\tau = \lambda^2/D = 1.7 \times 10^{-4}$~s under these conditions.
Using our observed negative net growth rate for $F = 0$ of $G = -\rhoeq^0/(\rho_0 \tau) = -0.00184$~ML~s$^{-1}$, this gives a value for the equilibrium adatom density of $\rhoeq^0 = 3.4 \times 10^{12}$~m$^{-2}$.
Using these estimates for $D$ and $\rhoeq^0$, the parameters obtained from the mixed kinetics fit imply kinetic coefficients of $\kappa_+^B = 0.74$~m~s$^{-1}$ and $\kappa_0^B = 1.3$~m~s$^{-1}$, and a step repulsion length of $\ell = 9 \times 10^{-10}$~m. 

Although it has not been possible using scanning-probe microscopy to observe the orientation difference of $\alpha$ and $\beta$ terraces on vicinal basal plane surfaces of HCP-type systems, our results show that this difference is robustly revealed by surface X-ray scattering.
\textit{In situ} X-ray measurements during growth can determine the fraction covered by each terrace, and thus distinguish the dynamics of $A$ and $B$ steps.
While the CTR calculations presented here are for wurtzite-structure GaN, this method applies to many other HCP-type systems with a $6_3$ screw axis, including other compound semiconductors, as well as one third of the crystalline elements and many more complex crystals.

The BCF model presented here makes predictions for the behavior of the $\alpha$ terrace fraction $f_\alpha$ at steady-state and during transients, in terms of surface properties such as the adatom diffusivity $D$ and step kinetic coefficients $\kappa_x^j$.
In particular, the steady-state fraction $\fass$ is predicted to depend only on the net growth rate $G = (F - \rhoeq^0 \tau)/\rho_0$, rather than individually on the deposition rate $F$ or the adatom lifetime $\tau$.
The positive or negative slope of $\fass(G)$ is determined by the sign of a combined kinetic parameter $\Kss$.

Our primary experimental result, the positive slope of $\fass(G)$, determines the basic nature of the adatom attachment kinetics at $A$ and $B$ steps for GaN (0001) OMVPE.
In general, this positive slope implies that $A$ steps have faster kinetics than $B$ steps, i.e. the attachment coefficients $\kappa_x^A$ are larger than the $\kappa_x^B$.
While a similar general shape of $\fass(G)$ is produced by many combinations of the parameters in the BCF model that have faster $A$ than $B$ step kinetics \cite{2020_Ju_PRB_BCF}, the best fit to our measurements is obtained in the specific limit of Eqs.~(\ref{eq:Dkm},\ref{eq:dfdtm}).
In this limit the $A$ step has much faster attachment kinetics than the $B$ step, with $\kappa_+^A >> \kappa_+^B$.
This limit also indicates that both $A$ and $B$ steps have standard positive Ehrlich-Schwoebel barriers, with adatom attachment from below significantly faster than from above for the same supersaturation, and that the $A$ step is non-transparent.
We find that $f_\alpha^0$ differs only slightly from the symmetrical value of 1/2.

In comparing the observed values of the step kinetic coefficients to predictions, the $5^\circ$ rotation of the step azimuth away from $[0 1 \overline{1} 0]$ is potentially important, since it determines the minimum average kink spacing on the steps to be $33$~\AA~ \cite{1951_Burton_PhilTransRS243_29}, {as discussed in Supplementary Discussion 2.}
We expect that this relatively small kink spacing will tend to produce higher predicted values of the attachment coefficients $\kappa_+^j$ and $\kappa_-^j$ and lower values of the transmission coefficients $\kappa_0^j$, since attachment occurs when adatoms at a step diffuse along it to find a kink, while transmission occurs when the adatom detaches from the step onto the opposite terrace before it finds a kink \cite{2007_Ranguelov_PRB75_245419}.
Because the transmitted adatom must traverse the barriers on both sides of the step independent of whether it arrives from above or below, $\kappa_0^j$ does not depend on the direction. 

Our conclusion that $A$ steps on GaN (0001) have higher attachment coefficients than $B$ steps agrees with the original qualitative prediction \cite{1999_Xie_PRL82_2749}, and motivates further quantitative theory development beyond that carried out to date.
In several previous studies \cite{2010_Zaluska-Kotur_JNoncrystSolids356_1935,2011_Zaluska-Kotur_JAP109_023515,2017_Chugh_ApplSurfSci422_1120,2017_Xu_JChemPhys146_144702,2013_Turski_JCrystGrowth367_115,2020_Akiyama_JCrystGrowth532_125410,2020_Akiyama_JJAP59_SGGK03,ohka2020effect}, various assumptions can lead to the opposite conclusion, as described below.
Both the predicted and observed step behavior can depend upon the chemical environment (e.g. OMVPE vs. MBE) and how it passivates the step edges.
For example, arguments regarding dangling bonds at steps \cite{1999_Xie_PRL82_2749,2013_Turski_JCrystGrowth367_115} depend on the effects of very high or low V/III ratios \cite{2017_Pristovsek_physstatsolb254_1600711} and the presence of NH$_3$ or H$_2$.
Likewise, KMC studies \cite{2011_Zaluska-Kotur_JAP109_023515,2010_Zaluska-Kotur_JNoncrystSolids356_1935,2017_Xu_JChemPhys146_144702,2017_Chugh_ApplSurfSci422_1120} typically make assumptions about bonding that determine the rates of atomic-scale processes at steps.
Detailed \textit{ab initio} predictions of kinetic barriers and adsorption energies at steps under MBE and OMVPE conditions \cite{2020_Akiyama_JCrystGrowth532_125410,2020_Akiyama_JJAP59_SGGK03,ohka2020effect} show that they depend strongly on the chemical environment and resulting surface reconstruction.
To date, these calculations have been carried out for the Ga(T4), NH(H3)+H(T1), and NH(H3)+NH$_2$(T1) reconstructions.
In future theoretical work, it would be useful to consider the specific step-edge structures associated with the 3H(T1) reconstruction found here and in recent calculations for the OMVPE environment \cite{2019_Kempisty_PRB100_085304}. 
Having an accurate model for atomic incorporation at steps can have important practical implications for advanced GaN devices, e.g. laser diodes and white LEDs, by allowing control of interface morphology and incorporation of alloying elements such as In \cite{2013_Turski_JCrystGrowth367_115,2016_Kaufmann_JCrystGrowth433_36}.

We have demonstrated this method using a micron-scale X-ray beam to satisfy the requirement that the illuminated surface region has a well-defined step azimuth.
With current synchrotron X-ray sources, it is convenient to increase the signal rate using a wide-energy-bandwidth pink beam.
The higher brightness synchrotron sources soon to come online worldwide will make it possible to perform this type of experiment with highly monochromatic beams, greatly increasing the in-plane resolution of the CTR measurements.


\section*{Methods}

\textbf{Calculated CTR intensities.}
Derivation of the CTR calculations for miscut surfaces with alternating terrace terminations is provided in a separate paper \cite{2020_Ju_PRB_CTR_arxiv}.
As described in Supplementary Discussion 1, in fits shown in Table~\ref{tab:table2}, and in recent predictions \cite{2019_Kempisty_PRB100_085304}, we expect that the GaN surface under OMVPE conditions has a 3H(T1) reconstruction, in which 3 of every 4 Ga atoms in top-layer sites shown in Fig.~\ref{fig:A_B} are bonded to an adsorbed hydrogen.
The calculations in Fig.~\ref{fig:CTR_miscut_3HT1}(a) include the effect of this reconstruction, with equal fractions of all reconstruction domains on both terraces.
Since no fractional-order diffraction peaks from long-range ordered reconstructions are observed in experiments, we expect that the domain structure has a short correlation length and all domains are present.
We use a surface roughness of $\sigma_R = 0.74$~\AA~to match the experimental fits described below.

\textbf{\textit{In situ} microbeam X-ray experiments.}
We performed \textit{in situ} measurements of the CTRs in the OMVPE environment at the Advanced Photon Source beamline 12ID-D \cite{2017_Ju_RSI88_035113}.
At an incidence angle of $2^\circ$, the 10~$\mu$m X-ray beam illuminated an area of $10 \times 300$~$\mu$m.
To obtain sufficient signal, we used a wide-bandwidth pink beam setup \cite{2018_Ju_JSyncRad25_1036,2019_Ju_NatPhys15_589}.
Details of the measurements and data analysis are given in Supplementary Methods 1.
Two types of measurements were performed.
We determined the steady-state terrace fractions $\fass$ under four different growth/evaporation conditions by scanning the detector along the $(0 1 \overline{1} L)$ and $(1 0 \overline{1} L)$ CTRs while continuously maintaining steady-state growth or evaporation.
We also observed the dynamics of $f_\alpha$ after an abrupt change in conditions.

Under the conditions studied, deposition is transport limited, with the deposition rate proportional to the supply of the Ga precursor (triethylgallium, TEGa), with a large excess of the N precursor (NH$_3$) constantly supplied.
We investigated conditions of zero deposition (no supply of TEGa) as well as deposition at a TEGa supply of 0.033~$\mu$mole~min$^{-1}$.
The NH$_3$ flow in both cases was 2.7~slpm or 0.12~mole~min$^{-1}$, and the total pressure was 267~mbar. 
The V/III ratio during deposition was thus $3.6 \times 10^6$.
For both of these conditions, we studied two carrier gas compositions: 50\% H$_2$ + 50\% N$_2$, and 0\% H$_2$ + 100\% N$_2$.
The addition of H$_2$ to the carrier gas enhances evaporation of GaN, so that the net growth rate (deposition rate minus evaporation rate) is slightly lower; at zero deposition rate, the net growth rate is negative.
We determined the net growth rate for all four conditions as described in Supplementary Methods 2.
These values are given in Table~\ref{tab:table1}.
Substrate temperatures were calibrated to within $\pm5$~K using laser interferometry from a standard sapphire substrate \cite{2017_Ju_RSI88_035113}.
While we used the same heater temperature for all conditions, the calibration indicates that the substrate temperature was slightly higher in 50\% H$_2$ (1080~K) than in 0\% H$_2$ (1073~K).
Based on an expected activation energy of $\sim 0.4$~eV for surface kinetics, this temperature difference produces a negligible (few percent) change in kinetic coefficients.

\textbf{Fits to steady-state CTRs.} For each growth condition, fits were performed of the calculated CTR intensities to the measured profiles of the $(0 1 \overline{1} L)$ and $(1 0 \overline{1} L)$ CTRs.
In addition to a single value of $f_\alpha$, parameters varied in the fit included a single surface roughness $\sigma_R$ and two intensity scale factors, one for each CTR.
We fit to $\log(I)$ with equal weighting of all points.
Fits were done using different possible surface reconstructions.
Supplementary Fig. 10 shows the calculated reconstruction phase diagram for the GaN (0001) surface in the OMVPE environment \cite{2012_Walkosz_PRB_85_033308}, as a function of Ga and NH$_3$ chemical potentials.
Based on the chemical potential values that correspond to our experimental conditions estimated in Supplementary Discussion 1, shown by the green rectangle, we considered the five reconstructions highlighted in Supplementary Fig. 10.
(The estimate for $\Delta \muGa$ has a large uncertainty because it depends on the nitrogen potential produced by decomposition of NH$_3$.)
Table~\ref{tab:table2} shows the values of steady-state terrace fraction $\fass$, surface roughness $\sigma_R$, and the goodness-of-fit parameter $\chi^2$ from fits to reflectivity at four conditions for each of five reconstructions.
The qualitative results for the variation of $\fass$ with growth condition are independent of which reconstruction is assumed: $\fass$ increases monotonically as the net growth rate $G$ increases.
The 3H(T1) reconstruction gives the best fit (minimum $\chi^2$) of the five potential reconstructions, for all four conditions.
This is consistent with recent results on GaN (0001) reconstructions in the OMVPE environment \cite{2019_Kempisty_PRB100_085304}, which found a phase stability region for the 3H(T1) structure even larger than that shown in Supplementary Fig. 10.
Fig.~\ref{fig:fit_nonspecular_exp_3HT1} compares the fits with the 3H(T1) reconstruction to the measured CTR intensities.

\begin{table}[h]
\caption{ \label{tab:table2} {Values of fit parameters for each of four OMVPE conditions.} We list the values of steady-state terrace fraction $\fass$, surface roughness $\sigma_R$, and the goodness-of-fit parameter $\chi^2$ from fits to reflectivity for each of 5 surface reconstructions.} 
\begin{ruledtabular}
\begin{tabular}{ c  l  l  l  l  }   
Growth & Reconstruction  &$\fass$ &$\sigma_R$ & $\chi^2$	\\
condition & & & (\AA) &  	\\
index  & & & & \\
\hline
& 3H(T1) & 0.111 & 0.75 & 106 \\
& Ga(T4) & 0.144 & 1.26 & 130 \\
1 & NH(H3)+H(T1)& 0.098 & 0.94 & 187\\
 & NH(H3)+NH$_2$(T1)& 0.106 & 0.88 & 200\\
& NH(H3)& 0.095& 0.91 & 167\\
\hline
& 3H(T1) & 0.461  & 0.93 & 57  \\
& Ga(T4) & 0.476 & 1.26 & 81 \\
2 & NH(H3)+H(T1)& 0.460 & 1.15 & 76\\
 & NH(H3)+NH$_2$(T1)& 0.460 & 1.11 & 67\\
& NH(H3)& 0.459& 1.13 & 99\\
\hline
& 3H(T1) & 0.811  & 0.85 & 118  \\
& Ga(T4) & 0.670 & 1.46  & 218 \\
3 & NH(H3)+H(T1)& 0.876 & 1.19 & 205\\
 & NH(H3)+NH$_2$(T1)& 0.869 & 1.15 & 248\\
& NH(H3)& 0.869& 1.16 & 168\\
\hline
& 3H(T1) & 0.868  & 0.47 & 80  \\
& Ga(T4) & 0.942 & 1.06  & 112 \\
4 & NH(H3)+H(T1)& 0.892 & 0.90 & 174\\
 & NH(H3)+NH$_2$(T1)& 0.879 & 0.85 & 220\\
& NH(H3)& 0.891& 0.87 & 135\\
\end{tabular}
\end{ruledtabular}
\end{table}

\textbf{Extracting terrace fraction dynamics.} We first convert the intensity evolution near $L = 1.6$ to reflectivity $R(t)$ by normalizing it to match the predicted change in reflectivity for the transition in $\fass$. 
We then invert the $R(f_\alpha)$ relation calculated for the experimental $L$ value, shown in Fig.~\ref{fig:CTR_miscut_3HT1}(b), to obtain $f_\alpha(t)$.
For simplicity we assume the surface roughness is not a function of growth condition, and use the average value of $\sigma_R = 0.74$~\AA~ from the 3H(T1) fits to calculate $R(f_\alpha)$.
If the surface roughness were to vary with condition by the amounts shown in Table~\ref{tab:table2}, this would explain only a small fraction of the observed changes (decrease of 7\% out of 40\% for the transition from condition 1 to 2, and increase of 16\% out of 350\% for the transition from condition 2 to 4).
The resulting $f_\alpha(t)$ are shown in Fig.~\ref{fig:relaxation_time}(b).
To extract a characteristic relaxation time for each transition, we interpolate the normalized the change $[f_\alpha(t) - f_\alpha(\infty)]/[f_\alpha(0) - f_\alpha(\infty)]$ to obtain the time at which it equals $1/e$.

\textbf{Fits of BCF theory to experimental results.} The general expressions for $\Kss(f_\alpha)$ and $\Kdyn(f_\alpha)$ \cite{2020_Ju_PRB_BCF} involve 9 unknown quantities ($D$, $\rhoeq^0 \ell^3$, $f_\alpha^0$, and the six $\kappa_x^j$) to be determined or constrained by the measurements.
This is a challenge because there are only 6 measured quantities (four steady-state $\alpha$ terrace fractions $\fass$ at different growth rates $G$, and two relaxation times for transitions in $G$.)
The best fit allowing all 9 parameters to vary was first determined by a numerical Levenberg-Marquardt nonlinear regression search algorithm. 
This minimized the goodness-of-fit parameter $\chi^2 \equiv \sum [(y_i - \yicalc)/\sigma_i]^2$, where the $y_i$ and $\sigma_i$ are the six measured quantities and their uncertainties.  
To estimate the uncertainties in the $\fass$, we multiplied those obtained in the fits to the 3H(T1) reconstruction by a factor of 4, to account for the uncertainties in the atomic coordinates used.
We estimated the uncertainty in the $\trel$ to be 10\%.

This initial fit found that the minimum $\chi^2$ occurs in the region of parameter space in which the $\kappa_+^A$ coefficient is large, while the $\kappa_-^A$, $\kappa_-^B$, and $\kappa_0^A$ coefficients approach zero.
In this case the limiting expressions Eqs.~(\ref{eq:Dkm}-\ref{eq:dfdtm}) for $\Kss(f_\alpha)$ and $\Kdyn(f_\alpha)$ involve only the four unknown quantities $D / \kappa_+^B$, $D / \kappa_0^B$, $D \rhoeq^0 \ell^3$, and $f_\alpha^0$. 
A final fit using Eqs.~(\ref{eq:Dkm}-\ref{eq:dfdtm}) allowing only these four parameters to vary gave the same solution as the 9 parameter fit.

\section*{Data availability}

Data supporting the findings of this study are available within the article and its Supplementary Information file and are available in electronic form from the corresponding author on reasonable request.

\begin{acknowledgments}
  Work supported by the U.S Department of Energy (DOE), Office of Science, Office of Basic Energy Sciences, Materials Science and Engineering Division.  Experiments were performed at the Advanced Photon Source beamline 12ID-D, a DOE Office of Science user facility operated by Argonne National Laboratory.
\end{acknowledgments}

\section*{Author Contributions}

All authors contributed to initial discussions motivating the experiments and analysis.
G.J., C.T., M.J.H., J.A.E., and G.B.S. developed the microbeam surface X-ray scattering method and carried out the measurements.
W.W. provided calculated atomic coordinates for the various surface reconstructions.
D.X., C.T., P.Z., and G.B.S. developed the CTR and BCF expressions used in the analysis.
G.J. and G.B.S. analyzed the results.
All coauthors contributed to drafting and editing the manuscript.

\section*{Competing Interests}

{The authors declare no competing interests.} 
  
\section*{Additional information}

Supplementary information is available for this paper at https://.
Correspondence and requests for materials or data should be addressed to G.B.S..

\end{document}


\title{Supplementary Information for ``\textit{In Situ} Microbeam Surface X-ray Scattering Reveals Alternating Step Kinetics During Crystal Growth''}

\author{Guangxu Ju}
    \email[correspondence to: ]{juguangxu@gmail.com}
	\altaffiliation[current address: ]{Lumileds Lighting Co., San Jose, CA 95131 USA.}
	\affiliation{Materials Science Division, Argonne National Laboratory, Lemont, IL 60439 USA}
\author{Dongwei Xu}
	\affiliation{Materials Science Division, Argonne National Laboratory, Lemont, IL 60439 USA}
	\affiliation{School of Energy and Power Engineering, Huazhong University of Science and Technology, Wuhan, Hubei 430074, China}
\author{Carol Thompson}
	\affiliation{Department of Physics, Northern Illinois University, DeKalb, IL 60115 USA}
\author{Matthew J. Highland}
	\affiliation{X-ray Science Division, Argonne National Laboratory, Lemont, IL 60439 USA}
\author{Jeffrey A. Eastman}
	\affiliation{Materials Science Division, Argonne National Laboratory, Lemont, IL 60439 USA}
\author{Weronika Walkosz}
	\affiliation{Department of Physics, Lake Forest College, Lake Forest, IL 60045 USA}
\author{Peter Zapol}
	\affiliation{Materials Science Division, Argonne National Laboratory, Lemont, IL 60439 USA}
\author{G. Brian Stephenson}
    \email[correspondence to: ]{stephenson@anl.gov}
	\affiliation{Materials Science Division, Argonne National Laboratory, Lemont, IL 60439 USA}

\date{\VERSION}

\maketitle

\begin{figure}
\includegraphics[width=\linewidth]{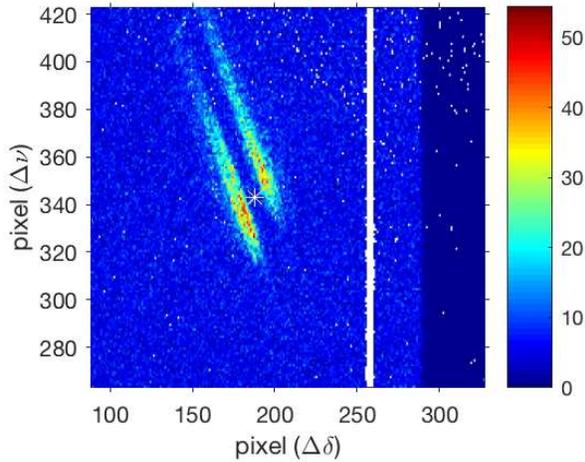}
\caption{\label{fig:CTR_angle_image} Detector image from $(0 1 \overline{1} L)$ scan for condition 4 ($0.033$~$\mu$mole~min$^{-1}$ TEGa, 0\% H$_2$), showing intensity maxima from Ewald sphere cutting through CTRs from the $(0 1 \overline{1} 1)$ and $(0 1 \overline{1} 2)$ Bragg peaks. Position of central pixel, marked by star, is $(0 1 \overline{1} L)$ with $L = 1.55$. Dark area on right is shadow of slits, and white vertical line is gap in pixels between detector chips. Pixels colored white are ignored due to excessive detector noise.}
\end{figure}

\begin{figure}
\includegraphics[width=\linewidth]{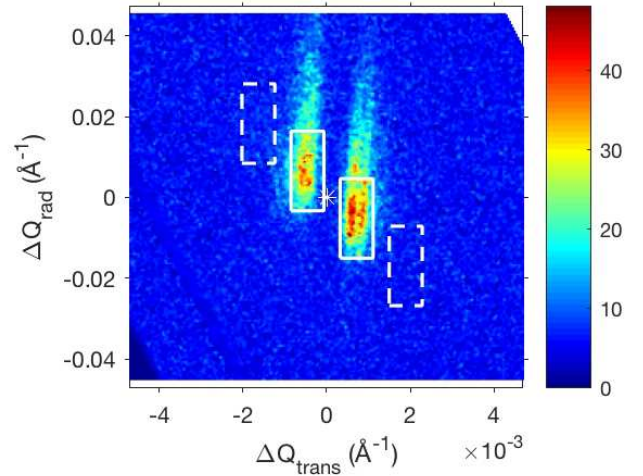}
\caption{\label{fig:CTR_qL_image} Cut through reciprocal space  at $L = Q_z c / 2 \pi = 1.55$, showing $(0 1 \overline{1} 1)$ and $(0 1 \overline{1} 2)$ CTRs. The in-plane $Q_x$ and $Q_y$ coordinates have been expressed as in-plane radial and transverse components $\Delta Q_\mathrm{rad}$ and $\Delta Q_\mathrm{trans}$ relative to the central pixel at position $(0 1 \overline{1} L)$. Solid and dashed rectangles show regions integrated to give CTR intensities and associated backgrounds, respectively.}
\end{figure}

\section*{Supplementary Methods 1:\\ X-ray Scattering}

To characterize the behavior of $A$ and $B$ steps on a GaN single crystal (0001) surfaces, we performed \textit{in situ} measurements of the CTRs during growth and evaporation in the OMVPE environment.  For details on the GaN single crystal substrate, see vendor website (GANKIBAN$^{TM}$ from SixPoint Materials, Inc., \url{spmaterials.com}).
We used a chamber and goniometer at the Advanced Photon Source beamline 12ID-D which were designed for \textit{in situ} surface X-ray scattering studies during growth \cite{2017_Ju_RSI88_035113}. For these experiments we used a modified chamber which had Be entrance and exit windows. A micron-scale X-ray beam illuminated a small surface area having a uniform step azimuth.
To obtain sufficient signal, we used a wide-bandwidth ``pink'' beam setup similar to that described previously \cite{2018_Ju_JSyncRad25_1036,2019_Ju_NatPhys15_589}.
The beam incident on the sample had a typical intensity of $1.4 \times 10^{12}$ photons per second at $E = 25.75$~keV, in a spot size of $10 \times 10$~$\mu$m.
At the $2^\circ$ incidence angle, this illuminated an area of $10 \times 300$~$\mu$m.
X-ray scattering patterns were recorded using a photon counting area detector with a GaAs sensor having 512 $\times$ 512 pixels, 55~$\mu$m pixel size, located $1.1$~m from the sample (Amsterdam Scientific Instruments LynX 1800). Slits downstream of the exit window block the scattering from the windows from reaching the region of the detector used to collect the CTR signal. 

To process the X-ray data from the area detector, raw images were first corrected for detector flatfield, eliminating pixels with excessive noise, and the signal was normalized to the incident intensity. 
Supplementary Fig.~\ref{fig:CTR_angle_image} shows a typical corrected detector image, with streaks from the $(0 1 \overline{1} 1)$ and $(0 1 \overline{1} 2)$ CTRs.
Because of the $\sim 1\%$ energy bandwidth of the pink beam \cite{2019_Ju_NatPhys15_589}, the CTRs are broadened radially as well as being extended in the surface normal $Q_z$ direction.
To convert the images along an $L$ scan to reciprocal space, the $Q_x Q_y Q_z$ coordinates of each pixel in each image were first calculated.
The out-of-plane coordinate $Q_z$ or $L$ varies across each image, following the Ewald sphere.
The in-plane coordinates $Q_x$ and $Q_y$ were converted to in-plane radial and transverse components $\Delta Q_\mathrm{rad}$ and $\Delta Q_\mathrm{trans}$ relative to the central position.
The intensities and $L$ values of each image were interpolated onto a fixed grid of $\Delta Q_\mathrm{rad}$ and $\Delta Q_\mathrm{trans}$.
We then interpolated the sequence of intensities from the scan at each $\Delta Q_\mathrm{rad}$ and $\Delta Q_\mathrm{trans}$ onto a grid of fixed $L$ values.

\begin{figure}
\includegraphics[width=0.9\linewidth]{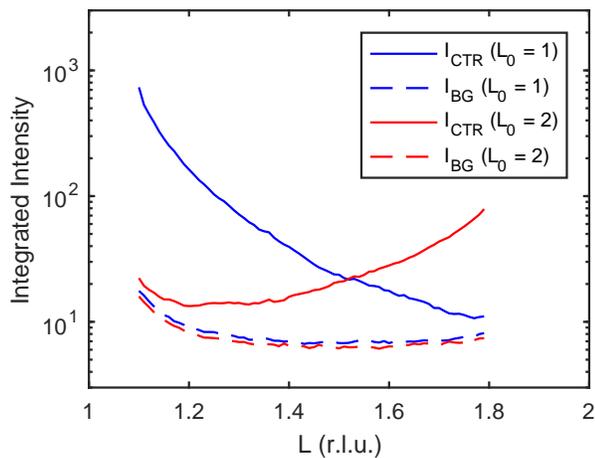}
\caption{\label{fig:CTR_backgrounds} Integrated total CTR and background intensities for the $(0 1 \overline{1} 1)$ and $(0 1 \overline{1} 2)$ CTRs, as a function of $L$ between $1.1$ and $1.8$, for condition 4 ($0.033$~$\mu$mole~min$^{-1}$ TEGa, 0\% H$_2$).}
\end{figure}

Supplementary Fig.~\ref{fig:CTR_qL_image} shows a typical cut through reciprocal space at fixed $L$.
The peaks from the $(0 1 \overline{1} 1)$ and $(0 1 \overline{1} 2)$ CTRs are conveniently separated in $\Delta Q_\mathrm{trans}$ because of the $5^\circ$ deviation of the step azimuth from $[0 1 \overline{1} 0]$; if the deviation had been zero, the peaks would have overlapped at $\Delta Q_\mathrm{trans} = 0$ because of the broadening in $\Delta Q_\mathrm{rad}$.
Regions of $\Delta Q_\mathrm{trans}$ and $\Delta Q_\mathrm{rad}$ surrounding each CTR were defined to integrate the total intensity, with positions that vary with $L$ to follow the CTRs.
Likewise adjacent regions were defined to integrate an equivalent volume of background scattering.
Such regions are shown as solid and dashed rectangles, respectively, in Supplementary Fig.~\ref{fig:CTR_qL_image}.
Supplementary Fig.~\ref{fig:CTR_backgrounds} shows the mean total CTR intensities and backgrounds in these regions as a function of $L$ for the scan between $L = 1.1$ and $L = 1.8$ for condition 4.
The net CTR intensity was calculated by subtracting the background from the total for that CTR. 
We ran scans from $L = 0.4$ to $L = 0.9$, $L = 1.1$ to $L = 1.8$, and $L = 2.15$ to $L = 2.6$ on the $(0 1 \overline{1} L)$ and $(1 0 \overline{1} L)$ CTRs, skipping over the Bragg peaks to avoiding having the high intensity strike the detector.
The $L$ range covered on each CTR varied depending upon the region covered by the detector in reciprocal space during the scan.

In order to determine whether exposure to the X-ray beam was affecting the OMVPE growth process, we periodically scanned the sample position while monitoring the CTR intensity.
For the conditions reported here, there was no indication that the spot which had been illuminated differed in any way from the neighboring regions.
During growth at higher temperatures (e.g. $1250$~K), we did observe local effects of the X-ray beam on the surface morphology.

\section*{Supplementary Methods 2:\\ Net Growth Rates}

Under the OMVPE conditions used, we observe that deposition of GaN is Ga transport limited (i.e. the deposition rate is proportional to the TEGa supply rate, nearly independent of $T$ and NH$_3$ supply), and the net growth rate has a negative offset at zero TEGa supply corresponding to an evaporation rate that depends on $T$ and the carrier gas composition (e.g. presence or absence of H$_2$).
To determine the deposition rate for the conditions used in the X-ray study, we used the deposition efficiency (deposition rate per TEGa supply rate) determined from previous studies of CTR oscillations during layer-by-layer growth \cite{2014_Perret_APL105_051602,2019_Ju_NatPhys15_589}.
We also measured the evaporation rates at two higher temperatures and both carrier gas compositions (0\% and 50\% H$_2$), and extrapolated them to the lower temperatures studied here.

\begin{figure}
\includegraphics[width=0.85\linewidth]{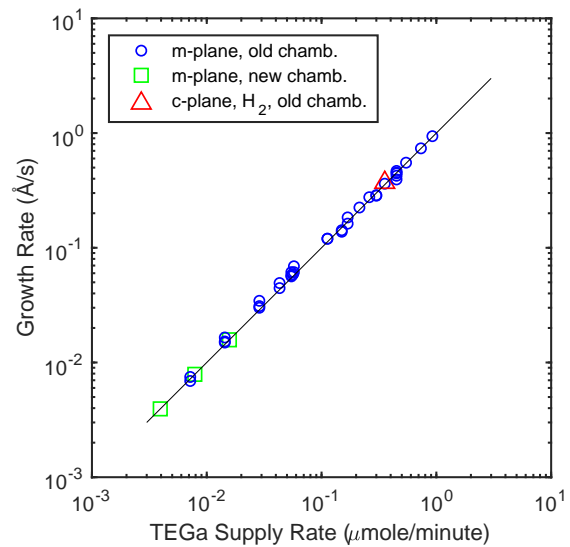}
\caption{Growth rate as a function of TEGa supply determined from CTR oscillations during layer-by-layer growth. Line is fit to new chamber data giving a deposition efficiency of 1.0 (\AA~s$^{-1}$)/($\mu$mole~min$^{-1}$). \label{fig:g0}}
\end{figure}

Supplementary Fig.~\ref{fig:g0} shows the growth rates measured from CTR oscillations during layer-by-layer growth as a function of TEGa supply \cite{2014_Perret_APL105_051602,2019_Ju_NatPhys15_589}.
In all cases the chamber flows were the same as in the X-ray study reported here (e.g. 2.7~slpm NH$_3$, 267~mbar total pressure).
Almost all data points are for growth on m-plane $(1 0 \overline{1} 0)$ GaN in N$_2$ carrier gas (0\% H$_2$), which exhibits layer-by-layer mode over a wide range of conditions.
Data are shown from both a previous growth chamber (``old'' chamber) \cite{1999_Stephenson_MRSBull24_21} and the current growth chamber (``new'' chamber) \cite{2017_Ju_RSI88_035113,2019_Ju_NatPhys15_589}.
The two chambers were designed to have the same flow geometry, and the growth behavior of both appear to be identical.
The data points from the previous chamber range in temperature from $848$~K to $1064$~K; the data points for the current chamber are for $867$~K.
The line shown is a fit to the data from the current chamber, which gives a deposition efficiency of 1.0 (\AA~s$^{-1}$)/($\mu$mole~min$^{-1}$).
One data point is shown for growth on c-plane (0001) GaN in 50\% N$_2$ + 50\% H$_2$ carrier gas at $900$~K; layer-by-layer growth was only observed on (0001) GaN under this condition.
It agrees with the m-plane data obtained in 0\% H$_2$ carrier, suggesting that the same deposition efficiency can be used for (0001) GaN in either 0\% or 50\% H$_2$ carrier gas.
We expect that there is negligible evaporation at $900$~K in either carrier gas.

To estimate the evaporation rate at the temperatures used in the X-ray study presented here (e.g. $1080$~K), we used laser interferometry to observe the change in thickness of an (0001) GaN film on a sapphire substrate \cite{2017_Ju_RSI88_035113,2005_Koleske_JCrystGrowth279_37}, under conditions of zero TEGa flow at higher $T$.
As the film thickness $d(t)$ changes during growth or evaporation, the back-scattered laser intensity $I(t)$ oscillates with time $t$ due to interference between light reflected from the film surface and the substrate/film interface, according to
\begin{equation}
    I(t) - I_\mathrm{min} = [I_\mathrm{max} - I_\mathrm{min}]
    \left ( \frac{1 + \cos[2 \pi d(t) / d_0]}{2} \right ),
\end{equation}
where $I_\mathrm{min}(t)$ and $I_\mathrm{max}(t)$ are the envelope of the minima and maxima, which can vary with time as film roughness changes, and the thickness oscillation period is $d_0 = \lambda/2n$, where $\lambda = 6330$~\AA ~is the wavelength of the light and $n$ is the refractive index of GaN.
This can be inverted to obtain the thickness evolution as
\begin{equation}
    d(t) = \left ( \frac{d_0}{2 \pi} \right ) \cos^{-1} \left ( \frac{2[I(t) - I_\mathrm{min}(t)]}{I_\mathrm{max}(t) - I_\mathrm{min}(t)} - 1 \right ). \label{eq:dt}
\end{equation}

\begin{figure}
\includegraphics[width=0.9\linewidth]{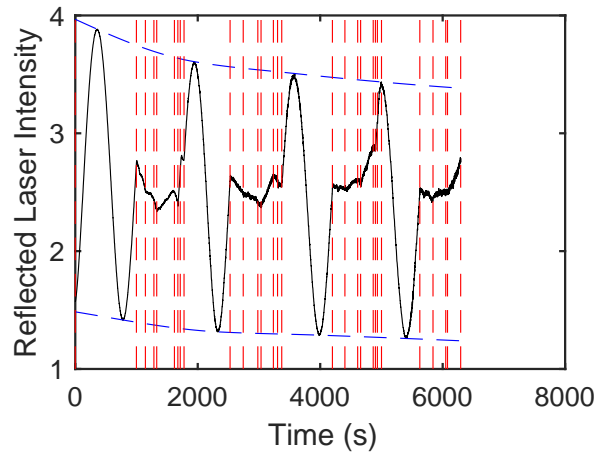}
\caption{Reflected laser signal during growth under various conditions. Vertical dashed lines show times at which conditions changed. \label{fig:g1}}
\end{figure}

\begin{figure}
\includegraphics[width=0.9\linewidth]{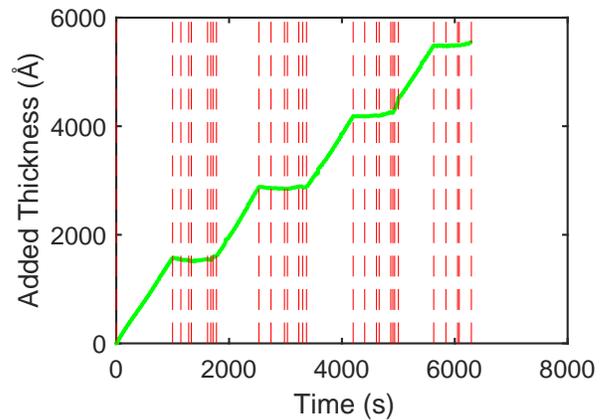}
\caption{Added thickness during growth under various conditions. Vertical dashed lines show times at which conditions changed. \label{fig:g2}}
\end{figure}

\begin{figure*}[t]
\includegraphics[width=0.75\linewidth]{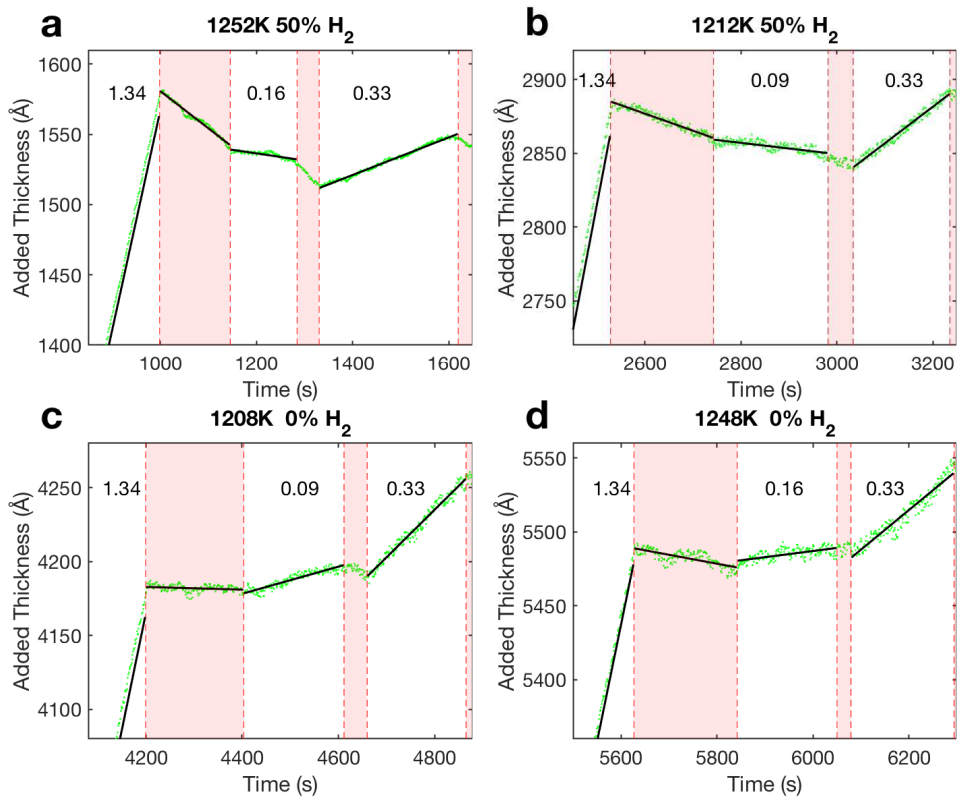}
\caption{Added thickness during growth under TEGa flows shown ($\mu$mole~min$^{-1}$) \textbf{a}. at T = 1252~K, 50\% H$_2$; \textbf{b}. at T = 1212~K, 50\% H$_2$; \textbf{c}. at T = 1208~K, 0\% H$_2$; \textbf{d}. at T = 1248~K, 0\% H$_2$. Vertical dashed lines show times at which conditions changed. Black lines show fits to extract net growth rates. TEGa flow was 0~$\mu$mole~min$^{-1}$ for the shaded regions.} \label{fig:g3}
\end{figure*}

Supplementary Fig.~\ref{fig:g1} shows the evolution of the laser signal with time during the experiment.
We began by growing a full oscillation at a high growth rate to obtain initial values for $I_\mathrm{min}$ and $I_\mathrm{max}$.
Once the signal had reach a value intermediate between these limits, where the phase of the oscillation is most accurately determined, we changed the TEGa flow $f_\mathrm{TEGa}$ to observe the net growth or evaporation rate at some fixed values of $f_\mathrm{TEGa}$.
Then we changed $T$ and/or the carrier gas concentration, and repeated the process starting with growing a full oscillation at a high rate.
The blue dashed curves in Supplementary Fig.~\ref{fig:g1} show the interpolated $I_\mathrm{min}(t)$ and $I_\mathrm{max}(t)$ envelopes.
Supplementary Fig.~\ref{fig:g2} shows the thickness change with time extracted with Supplementary Eq.~(\ref{eq:dt}), using a value of $d_0 = 1302$~\AA ~corresponding to $n = 2.431$ \cite{1986_Tapping_JOSAA3_610,1977_Touloulian_TPRCseries13}.
Supplementary Fig.~\ref{fig:g3} shows expanded regions of the thickness evolution, where we varied $f_\mathrm{TEGa}$ for specific $T$ and H$_2$ fractions.
The solid lines show linear fits to extract the net growth rate in  \AA~s$^{-1}$ at each value of $f_\mathrm{TEGa}$.
This is equal to $Gc/2$, where $G$ is the growth rate in ML~s$^{-1}$ and $c/2$ is the thickness of a monolayer (ML).

\begin{figure}
\includegraphics[width=0.8\linewidth]{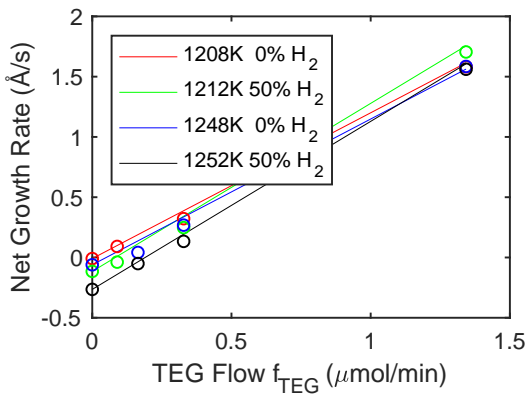}
\caption{Symbols show extracted net growth rates vs. TEGa flow for various $T$ and carrier gas compositions. \label{fig:g7}}
\end{figure}

\begin{figure}
\includegraphics[width=0.85\linewidth]{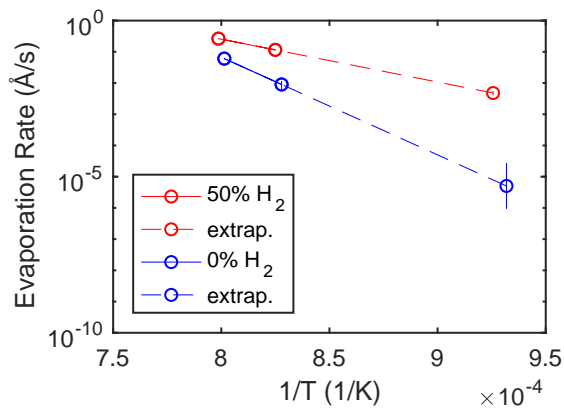}
\caption{Evaporation rate at zero TEGa flow as a function of $T$, with and without H$_2$, with extrapolation to lower $T$. \label{fig:g8}}
\end{figure}

The extracted values of $Gc/2$ are given in Supplementary Table~\ref{tab:table3} and plotted in Supplementary Fig.~\ref{fig:g7} as a function of $f_\mathrm{TEGa}$.
We observe that $Gc/2$ becomes negative at $f_\mathrm{TEGa} = 0$ due to evaporation, and that evaporation is more rapid at higher $T$ and when H$_2$ is present in the carrier gas.
These evaporation rates in 50\% H$_2$ are consistent with the rate of $\rho_0 G = -4.2 \times 10^{18}$~m$^{-2}$s$^{-1}$ or $Gc/2 = -0.97$~\AA~s$^{-1}$ given in the literature \cite{2001_Koleske_JCrystGrowth223_466} at a higher $T = 1300$~K with H$_2$ and NH$_3$ at a total pressure of $267$~mbar.
Also given in Supplementary Table~\ref{tab:table3} are the deposition efficiencies $d(Gc/2)/df_\mathrm{TEGa}$ obtained from linear fits to $Gc/2$ at the four values of $f_\mathrm{TEGa}$ for each $T$ and H$_2$ fraction.
The values are all similar to but slightly higher than the value of $d(Gc/2)/df_\mathrm{TEGa} = 1.0$~(\AA~s$^{-1}$)/($\mu$mole~min$^{-1}$) that we have observed from growth oscillations during layer-by-layer growth at lower $T$, described above \cite{2019_Ju_NatPhys15_589,2014_Perret_APL105_051602}.
The efficiency seems to be slightly larger for 50\% H$_2$ compared with 0\% H$_2$.
This may indicate that the deposition efficiency can vary somewhat as the flow and diffusion fields vary in the chamber with $T$ or carrier gas composition. 

To obtain the evaporation rate at the lower $T$ used in the X-ray experiments, we extrapolated the values at $f_\mathrm{TEGa} = 0$ for 50\% H$_2$ or 0\% H$_2$ assuming Arrhenius behavior of the evaporation rate, as shown in Supplementary Fig.~\ref{fig:g8}.
The fitted activation energies are $2.7 \pm 0.1$ and $6.2 \pm 1.2$~eV in 50\% and 0\% H$_2$, respectively.
We obtain evaporation rates of $4.8 \pm 0.8 \times 10^{-3} $~\AA~s$^{-1}$ at $T = 1080$~K with 50\% H$_2$, and $5 \times 10^{-6}$~\AA~s$^{-1}$ (with error limits of a factor of 5) at $T = 1073$~K with 0\% H$_2$.
We have used these evaporation rates, as well as the low-temperature deposition efficiency of $1.0$~(\AA~s$^{-1}$)/($\mu$mole~min$^{-1}$) and the TEGa flow rates of $0$ or $0.033$~$\mu$mole~min$^{-1}$, to calculate the net growth rates given in Table I in the main text for the 4 conditions studied.

\begin{table} 
\caption{ \label{tab:table3} Values of net growth rate $Gc/2$ extracted from laser interferometry measurements. We list the values for two temperatures and for carrier gas with and without H$_2$, as a function of TEGa flow $f_\mathrm{TEGa}$. Also shown is fitted $d(Gc/2)/df_\mathrm{TEGa}$ for each $T$ and carrier gas.} 
\begin{ruledtabular}
\begin{tabular}{ c | c | c || c | c }   
$T$ & H$_2$ & $f_\mathrm{TEGa}$ & $Gc/2$  & $d(Gc/2)/df_\mathrm{TEGa}$	\\
(K) & in& ($\mu$mole & (\AA~s$^{-1}$) & (\AA~s$^{-1}$)/ \\
  & carr. & min$^{-1}$) & & ($\mu$mole~min$^{-1}$) \\
\hline
 1208 & 0\%& 0.00 & $-0.009 \pm 0.003$ & $1.19 \pm 0.03$ \\
& & 0.09 & $0.092 \pm 0.003$ & \\
& & 0.33 & $0.322 \pm 0.004$ & \\
& & 1.34 & $1.582 \pm 0.002$ & \\
\hline
 1212 & 50\%& 0.00 & $-0.115 \pm 0.002$ & $1.38 \pm 0.05$ \\
& & 0.09 & $-0.038 \pm 0.002$ & \\
& & 0.33 & $0.248 \pm 0.003$ & \\
& & 1.34 & $1.705 \pm 0.002$ & \\
\hline
 1248 & 0\%& 0.00 & $-0.061 \pm 0.004$ & $1.27 \pm 0.05$ \\
& & 0.16 & $0.042 \pm 0.004$ & \\
& & 0.33 & $0.268 \pm 0.005$ & \\
& & 1.34 & $1.584 \pm 0.002$ & \\
\hline
 1252 & 50\%& 0.00 & $-0.265 \pm 0.004$ & $1.40 \pm 0.03$ \\
& & 0.16 & $-0.050 \pm 0.003$ & \\
& & 0.33 & $0.134 \pm 0.001$ & \\
& & 1.34 & $1.562 \pm 0.001$ & \\
\end{tabular}
\end{ruledtabular}
\end{table}

\section*{Supplementary Discussion 1:\\ Chemical potentials in OMVPE}

To calculate the CTR intensities to fit to the experimental profiles, we need to include the effect of reconstruction of the surface.
The relaxed atomic structures and free energies of various surface reconstructions for GaN (0001) in the OMVPE environment containing NH$_3$ and H$_2$ have been calculated \cite{2002_VandeWalle_PRL88_066103, 2012_Walkosz_PRB_85_033308}, leading to a phase diagram that can be expressed in terms of the chemical potentials of Ga and NH$_3$ \cite{2002_VandeWalle_JVacSciTechnolB20_1640,2012_Walkosz_PRB_85_033308}.
In this section we estimate these chemical potentials from the conditions in our experiments, to locate the appropriate region of the phase diagram and identify the predicted reconstructions in this region.

\begin{figure}
\includegraphics[width=0.9\linewidth]{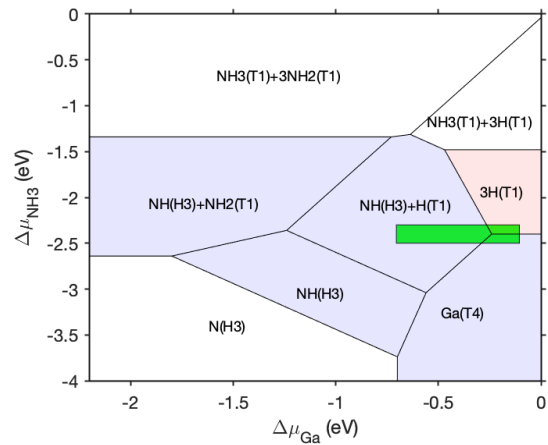}
\caption{\label{fig:recon_phase} Surface reconstruction phase diagram for GaN (0001), calculated in \cite{2012_Walkosz_PRB_85_033308}. Green rectangle shows estimated position of our experimental conditions. Five shaded reconstructions near these conditions were considered in fits shown in Table II. The 3H(T1) reconstruction gives the
best fit to all conditions.}
\end{figure}

Supplementary Fig.~\ref{fig:recon_phase} shows the predicted surface phase diagram \cite{2012_Walkosz_PRB_85_033308}. 
The vertical axis is the chemical potential of NH$_3$ relative to its value at $T = 0$~K.
This can be expressed as
\begin{align}
    \Delta \mu_\mathrm{NH_3}(T) &\equiv \mu_\mathrm{NH_3}(T) - \mu_\mathrm{NH_3}(0) \nonumber \\
    &= G_\mathrm{NH_3}^{\circ}(T) - G_\mathrm{NH_3}^{\circ}(0) + kT \log p_\mathrm{NH_3},
    \label{eq:DmuNH3}
\end{align}
where $G_\mathrm{NH_3}^{\circ}$ is the free energy of NH$_3$ gas at a pressure of 1~bar obtained from thermochemical tables \cite{1998_Chase_JPCRDMono9_NIST-JANAF},
and $p_\mathrm{NH_3}$ is the partial pressure of NH$_3$ in the experiment.
These can be evaluated at the experimental conditions.
For $T = 1073$~K, the tables give $G_\mathrm{NH_3}^{\circ}(T) - G_\mathrm{NH_3}^{\circ}(0) = -2.1$~eV.
Thus for $p_\mathrm{NH_3} = 0.04$~bar, one obtains 
$\Delta \mu_\mathrm{NH_3}(T) = -2.4$~eV.

The horizontal axis in Supplementary Fig.~\ref{fig:recon_phase} is the chemical potential of Ga relative elemental liquid Ga.
This can be related to the activity of N$_2$ using
\begin{align}
    \Delta \mu_\mathrm{Ga} &\equiv \mu_\mathrm{Ga}(T) - \mu_\mathrm{Ga}^\mathrm{liq}(T) \nonumber \\ 
    &= \Delta G_{f}^\mathrm{GaN}(T) - 0.5 kT \log a_\mathrm{N_2},
    \label{eq:DmuGa}
\end{align}
where $\Delta G_{f}^\mathrm{GaN}$ is the free energy of formation of GaN from liquid Ga and N$_2$ gas at 1~bar, and $a_\mathrm{N_2}$ is the activity (effective partial pressure) of N$_2$.

In OMVPE, a chemically active precursor such as ammonia is typically used to provide the high nitrogen activity required to grow group III nitrides.
The need for this can be seen in Supplementary Fig.~\ref{fig:DGf}, which shows the free energies of the reactions to form GaN and InN from the condensed metallic elements and either vapor N$_2$ or vapor NH$_3$ at 1~bar \cite{1998_Chase_JPCRDMono9_NIST-JANAF,1996_Ambacher_JVSTB14_3532}.
At typical temperatures used for growth of high quality single crystal films at high rates (e.g. 1000~K for InN, 1300~K for GaN), the formation energy from N$_2$ is positive, indicating that the nitride is not stable and cannot be grown from N$_2$ at 1~bar.
In contrast, the formation energies of the nitrides (plus H$_2$ at 1~bar) from the metals and NH$_3$ are negative at all relevant growth temperatures,
indicating that growth from 1~bar of NH$_3$ is possible.

\begin{figure}
\includegraphics[width=3.0in]{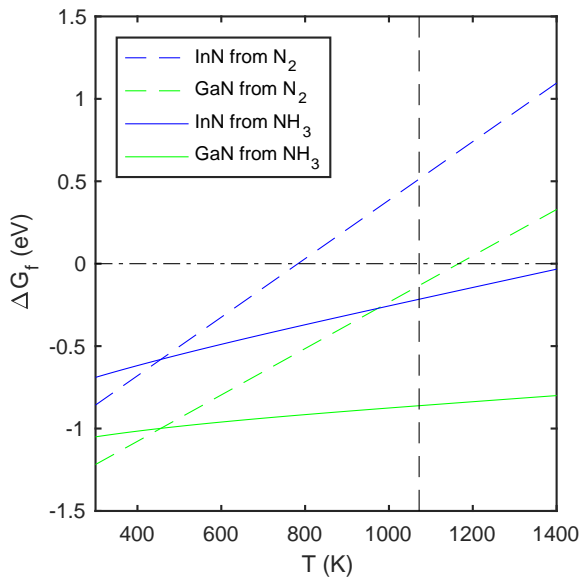}
\caption{\label{fig:DGf} Free energy of formation as a function of temperature of InN and GaN from the liquid metals and either vapor N$_2$ or NH$_3$ at 1~bar \cite{1998_Chase_JPCRDMono9_NIST-JANAF,1996_Ambacher_JVSTB14_3532}. In the case of NH$_3$, this includes formation of H$_2$ at 1~bar. The vertical dashed black line corresponds to the $T$ used in the current experiment.}
\end{figure}

However, actual OMVPE conditions do not correspond with equilibrium, because the very high partial pressures of N$_2$ and/or H$_2$ that would correspond to equilibrium with NH$_3$ at these temperatures are not allowed to accumulate.
Thus, while formation of InN and GaN from NH$_3$ is energetically favored under OMVPE conditions, decomposition of these nitrides into N$_2$ is also energetically favored.
This metastability is manifested in the oscillatory growth and decomposition of InN that has been observed~\cite{2008_Jiang_PRL101_086102}.
Thus the kinetics of the reaction steps that determine the nitrogen activity at the growth surface are critical to understanding and controlling OMPVE growth of metastable nitrides.

In previous work we have measured the trimethylindium (TMI) partial pressures required to condense InN and elemental In onto GaN (0001) \cite{2008_Jiang_PRL101_086102}.
They can be analyzed to give experimentally determined values for the effective surface nitrogen activity arising from NH$_3$ under OMVPE conditions.
The experiments were carried out using a very similar growth chamber~\cite{1999_Stephenson_MRSBull24_21} as that used for the {\it in situ} X-ray studies described below,
using the same a total pressure of 0.267~bar, and the same NH$_3$ and carrier flows (2.7 standard liters per minute (slpm) NH$_3$ and 1.1~slpm N$_2$ in the group V channel, 0.9~slpm N$_2$ carrier gas for TMI in the group III channel). 
We have performed chamber flow modeling to calculate the equivalent TMI and NH$_3$ partial pressures $p_\mathrm{TMI}$ and $p_\mathrm{NH_3}$ above the center of the substrate surface as a function of inlet flows.
At typical growth temperatures, an inlet flow of 0.184~$\mu$mole~min$^{-1}$ TMI corresponds to $p_\mathrm{TMI} = 1.22 \times 10^{-6}$~bar, and 
an inlet flow of 2.7 slpm NH$_3$ corresponds to $p_\mathrm{NH_3} = 0.040$~bar.

\begin{table} 
\caption{ \label{tab:aN2} Evaluation of N$_2$ activity and $\Delta \mu_\mathrm{Ga}$ at the GaN surface under OMVPE conditions. Formation energies of GaN and InN are from elements at standard conditions. TMI pressures at InN and In condensation boundaries are for $p_\mathrm{NH_3} = 0.04$~bar. Calculated $a_\mathrm{N_2}$ and $\Delta \mu_\mathrm{Ga}$ are thus also for $p_\mathrm{NH_3} = 0.04$~bar.}
\begin{ruledtabular}
\begin{tabular}{ c || c }
Quantity & Value as $f(T)$~(K)\\
 & (eV) \\
\hline
$\Delta G_f^\mathrm{GaN}$ \cite{1996_Ambacher_JVSTB14_3532} & $-1.64 + 1.41 \times 10^{-3} T$ \\
$\Delta G_f^\mathrm{InN}$ \cite{1996_Ambacher_JVSTB14_3532} & $-1.39 + 1.78 \times 10^{-3} T$ \\
$kT \log p_\mathrm{TMI}^\mathrm{InN}$ \cite{2008_Jiang_PRL101_086102} & $-1.309 + 0.88 \times 10^{-3} T$ \\
$kT \log p_\mathrm{TMI}^\mathrm{In}$ \cite{2008_Jiang_PRL101_086102} & $-3.843 + 3.47 \times 10^{-3} T$ \\
\hline
$kT \log a_\mathrm{In}$ & $-2.534 + 2.59 \times 10^{-3} T$ \\
$= kT \log p_\mathrm{TMI}^\mathrm{InN} - kT \log p_\mathrm{TMI}^\mathrm{In}$ & \\
\hline
$kT \log a_\mathrm{N_2}$ & $2.288 - 1.63 \times 10^{-3} T$ \\
$= 2(\Delta G_f^\mathrm{InN} - kT \log a_\mathrm{In})$ & \\
\hline
$\Delta \mu_\mathrm{Ga}$ & $-2.784 + 2.225 \times 10^{-3} T$ \\
$= \Delta G_f^\mathrm{GaN} - 0.5 kT \log a_\mathrm{N_2}$ & \\
\end{tabular}
\end{ruledtabular}
\end{table}

\begin{figure}
\includegraphics[width=3.0in]{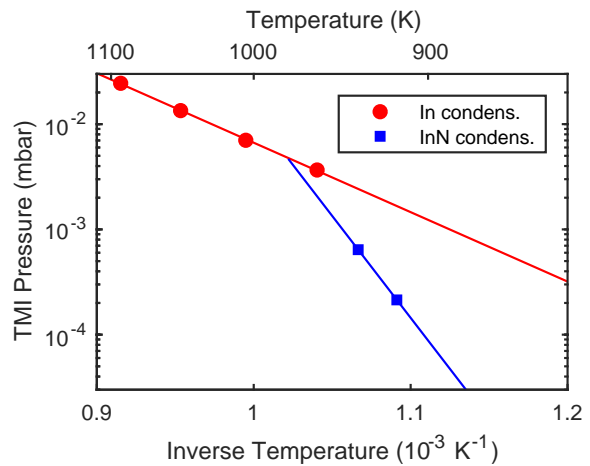}
\caption{\label{fig:InN} Observed phase boundaries for condensation onto GaN (0001) of relaxed epitaxial InN (blue squares) and liquid elemental In (red circles) at $p_\mathrm{NH_3} = 0.040$~bar \cite{2008_Jiang_PRL101_086102}.}
\end{figure}

Supplementary Fig.~\ref{fig:InN} shows the $p_\mathrm{TMI}$-$T$ boundaries determined by {\it in situ} X-ray fluorescence and diffraction measurements for initial condensation of elemental In liquid or crystalline InN onto a GaN (0001) surface at $p_\mathrm{NH_3} = 0.040$~bar \cite{2008_Jiang_PRL101_086102}. 
At TMI partial pressures above the boundaries shown, the condensed phases nucleate and grow on the surface;
at lower $p_\mathrm{TMI}$, the condensed phases evaporate. 
The InN and In condensation boundaries intersect at 979~K. 

A relationship between the nitrogen and indium activities at the InN condensation boundary can be obtained from the equilibrium
\begin{equation}
\mathrm{In}_{vap} + \frac12 \mathrm{N}_2 \leftrightarrow \mathrm{InN}_{sol},
\end{equation}
which gives the chemical potential $\mu_i$ expression
\begin{equation}
\mu_\mathrm{In} + \frac12 \mu_\mathrm{N_2} = \mu_\mathrm{InN},
\end{equation}
and the activity $a_i$ expression
\begin{equation}
kT \log a_\mathrm{In} + \frac12 kT \log a_\mathrm{N_2} = \Delta G_f^\mathrm{InN}(T),
\label{eq:aIn}
\end{equation}
where $\Delta G_f^\mathrm{InN}(T)$ is the formation energy of InN from liquid In and N$_2$ at 1~bar shown in Supplementary Fig.~\ref{fig:InN}. 
We assume that the activity of In relative to liquid In at the InN boundary is equal to the ratio $a_\mathrm{In} = p_\mathrm{TMI}^\mathrm{InN} / p_\mathrm{TMI}^\mathrm{In}$, giving
\begin{equation}
kT \log a_\mathrm{In} = kT \log p_\mathrm{TMI}^\mathrm{InN} - kT \log p_\mathrm{TMI}^\mathrm{In}
\end{equation}
at the experimental condition, $p_\mathrm{NH_3} = 0.040$~bar. 
Supplementary Eq.~(\ref{eq:aIn}) can then be used to obtain the nitrogen activity relative to 1~bar (i.e. effective partial pressure of N$_2$ in bar) for $p_\mathrm{NH_3} = 0.040$~bar. 

Supplementary Table~\ref{tab:aN2} summarizes the calculations to obtain the nitrogen activity and $\Delta \mu_\mathrm{Ga}$ under our OMVPE conditions.
The value of $k T \log a_\mathrm{N_2} = 0.55$~eV at the experimental temperature $T = 1073$~K gives the horizontal coordinate on the phase diagram from Supplementary Eq.~(\ref{eq:DmuGa}) as $\Delta \mu_\mathrm{Ga} = -0.40$~eV.
The value of $k T \log p_\mathrm{NH_3} = -0.30$~eV at the experimental temperature $T = 1073$~K gives the vertical coordinate on the phase diagram from Supplementary Eq.~(\ref{eq:DmuNH3}) as $\Delta \mu_\mathrm{NH_3} = -2.40$~eV.
This position is shown on the predicted surface phase diagram, Supplementary Fig.~\ref{fig:recon_phase}, with a rectangle representing the relatively large uncertainty in $\Delta \mu_\mathrm{Ga}$.

A recent study of reconstructions on GaN (0001) in the OMVPE environment \cite{2019_Kempisty_PRB100_085304}
included the effects of additional entropy associated with adsorbed species, which leads to a phase diagram that varies somewhat with temperature, even when expressed in chemical potential coordinates.
These effects tend to stabilize reconstructions with H adsorbates at higher $T$, giving a larger region of phase stability for the 3H(T1) reconstruction than shown in Supplementary Fig.~\ref{fig:recon_phase}.
This is consistent with our finding that the 3H(T1) reconstruction agrees best with the experimental CTRs for all conditions studied.

\section*{Supplementary Discussion 2:\\ Kink Density}

The density of kinks on a step is determined by two terms: the minimum density that is geometrically required to give the average direction of the step, and the additional thermally generated kink pairs \cite{1951_Burton_PhilTransRS243_29}.
This can be analyzed in terms of the probabilities $n_+$ and $n_-$ for positive or negative kinks to occur at each lattice site on the step.
For a close-packed surface with lattice parameter $a$, the geometrical requirement gives
\begin{equation}
\theta \equiv n_+ - n_- = 2 / (\sqrt{3}/ \tan \phi + 1), 
\end{equation}
where $\phi$ is the angle of the step with respect to the atomic rows in the $[2 \overline{1} \overline{1} 0]$ type directions.
The probabilities must also satisfy
\begin{equation}
    n_+ n_- = \exp(-2 w / kT),
\end{equation}
where $2w$ is the energy cost to generate a kink pair, and we assume the kink probabilities are much smaller than unity.

For our sample with $\phi = 5^\circ$ and $a = 3.2$~\AA, the geometrically required maximum average kink spacing is $a / \theta = 33$~\AA. 
If we estimate the kink pair energy as $2 w = W / 6$ where $W = 3.38$~eV is the bulk binding energy per molecule for GaN \cite{2017_Xu_JChemPhys146_144702}, this gives an average kink spacing of $a / (\theta + 2 n_-) = 24$~\AA~at $T = 1073$~K. 


\begin{acknowledgments}
  Work supported by the U.S Department of Energy (DOE), Office of Science, Office of Basic Energy Sciences, Materials Science and Engineering Division.  Experiments were performed at the Advanced Photon Source beamline 12ID-D, a DOE Office of Science user facility operated by Argonne National Laboratory.
\end{acknowledgments}


\bibliography{bibliography/2020_GJu_ABsteps_full_shortJnames}
